\documentclass[aps,prb,twocolumn]{revtex4-2}
\usepackage{graphicx}
\usepackage{amsmath}
\usepackage{amssymb}
\usepackage{commath}
\usepackage{braket}
\usepackage{hyperref}
\usepackage{subfigure}
\usepackage[usenames,dvipsnames]{color}

\usepackage[mathscr]{euscript}
\usepackage{bm}
\usepackage{dcolumn}
\usepackage{tikz,pgfplots}
\usetikzlibrary{decorations.pathmorphing,patterns,positioning}
\usetikzlibrary{tikzmark}
\usetikzlibrary{shapes,shadows,arrows}

\newcommand\beq{\begin{equation}}
\newcommand\eeq{\end{equation}}
\newcommand\bea{\begin{eqnarray}}
\newcommand\eea{\end{eqnarray}}

\newcommand\non{\nonumber}

\newcommand{\dket}[1]{|#1\rangle\rangle}

\usepackage{comment}
\usepackage[normalem]{ulem}

\pgfplotsset{compat=1.18}

\begin{document}
	
\title{Kinetically constrained models constructed from dissipative quantum dynamics} 
	
\author{ Somnath Maity$^1$ and 
		 Ryusuke Hamazaki$^{1,2}$}
\affiliation{$^1$Nonequilibrium Quantum Statistical Mechanics RIKEN Hakubi Research Team, RIKEN Cluster for Pioneering Research (CPR),
	2-1 Hirosawa, Wako, Saitama 351-0198, Japan \\
		$^2$RIKEN Interdisciplinary Theoretical and Mathematical Sciences Program (iTHEMS), 2-1 Hirosawa, Wako, Saitama 351-0198, Japan}
	
\date{\today}
	 
	\begin{abstract}
We propose a construction of kinetically constrained models using the Markovian quantum dynamics under strong dissipation. 
{Engineering the Gorini-Kossakowski-Sudarshan-Lindblad (GKSL) equation through classical noise, we show that strong dissipation leads to the emergent decoherence-free subspaces, within which constrained quantum many-body unitary dynamics can take place.}
We argue that the unitary dynamics constructed by the GKSL dynamics is more tightly constrained than that constructed by the strongly interacting Hamiltonian, where the interactions have the same form with the GKSL jump operators.
As an example, we demonstrate that a one-dimensional spin system with two-site dissipation leads to the kinetically constrained ``PXQ" model, which exhibits the free domain-wall motion with an additional frozen-block structure.
Under a uniform magnetic field, the PXQ model shows the domain-wall localization, similar to the Wannier-Stark localization.
We then couple two PXQ chains with the magnetic field by an inter-chain interaction.
We discover that, while localization of the domain walls persists despite the interactions for typical parameter regimes,  
a non-trivial partial delocalization appears for a certain parameter line.
\end{abstract}
	
\maketitle
	
\section{Introduction}
\label{sec_intro}
In the past two decades, a great effort has been devoted to understanding thermalization dynamics of isolated quantum many-body systems~\cite{polkovnikov11,luca16,gogolin16}, owing to remarkable advancements in experimental techniques~\cite{kinoshita06,gring12,schreiber15,hild14,kaufman16,tang18}. With these developments, it is now widely recognized that the eigenstate thermalization hypothesis (ETH)~\cite{deutsch91,srednicki,srednicki99,rigol08} provides the underlying mechanism for isolated quantum systems to thermalize. It conjectures that mid-spectrum eigenstates of a generic closed quantum system are intrinsically thermal as far as expectation values of local observables are concerned~\cite{luca16,kim14,Beugeling14,khodja15,Mondaini16,yoshizawa18, Jansen19,sugimoto21,Sugimoto22}. In the past decade, there has been a heightened interest in models violating the ETH. The search for such nonergodic quantum systems is crucial as they preserve quantum information in the system without thermalization.

Recently, there has been an emphasis on the study of systems displaying nonergodic dynamics without quenched disorder (unlike many-body localization~\cite{prozen08,pal10,nandkishore15,schreiber15,abanin19}), especially the so-called kinetically constrained models (KCMs).
Originally introduced to describe the behavior of classical glasses~\cite{ritort03,toninelli04,garrhan07}, the quantum version of the KCMs have recently attracted significant interest in this context~\cite{olmos14,lan18,singh21,scherg21,tortora22}.  
One such example of kinetically constrained models is the celebrated PXP model~\cite{turner18nature,turner18,khemani19}.
Here, nearest-neighbor excitations are energetically prohibited due to strong coupling 
in a chain of ultracold Rydberg atoms, and a long-lived coherent oscillation is observed for certain initial states~\cite{bernien17}.
This is because of some athermal eigenstates within the otherwise thermal many-body spectrum, which are dubbed as quantum many-body scars~\cite{serbyn21,moudgalya22,chandran23,moudgalya18,moudgalya18_1,schecter19,mark20,desaules21}.
As another example, the phenomenon called {Hilbert space fragmentation (HSF)} is also known to occur in certain KCMs~\cite{sala20,khemani20,rakovszky20,Gorshkov20,Moudgalya22_prx,Yoshinaga22,Hart22}.
The Hamiltonian of such systems is fragmented into an exponentially large number of  dynamically disconnected blocks in the computational basis and hence large parts of the Hilbert space are inaccessible to a particular initial state.
Furthermore, the HSF can also be used to describe the nonergodic behavior observed in systems with a tilted potential~\cite{scherg21,morong21,kohlert23} referred to as the Wannier-Stark many-body localization~\cite{schulz19,Nieuwenburg19}. 
Crucially, these KCMs are typically obtained as an effective description of the systems under strong interaction compared to the kinetic terms.

For  robust storage of quantum information, another important factor for modern quantum technologies is dissipation induced by an external environment. 
Indeed, many of the above phenomena rely on such delicate properties of the system that even a weak coupling with the external environment could have adverse implications~\cite{Nandkishore14,Medvedyeva16,Fischer16,levi16,Luschen17,Wu19,wybo20,Fujimoto22}. On the other hand, it has also been realized that dissipation engineering, the method of controlling interactions between the system and the environment~\cite{Poyatos96,Harrington22}, can be used for various quantum tasks, such as 
preparation of nontrivial nonequilibrium states~\cite{diehl08,buca19,Diehl11,Budich15,Bardyn12}, quantum computation~\cite{verstraete09}, and quantum error correction~\cite{Leghtas13}. The interplay between Hamiltonian interaction and coupling to the environment even leads to a new kind of phase transition and criticality~\cite{kessler12,minganti18,Haga23,rossini21}, represented by dissipative time crystals~\cite{Gong18,buca19,Iemini18,Gambetta19,kessler21,Kongkhambut22,taheri22}.
Furthermore, coherent quantum dynamics are also simulated in the steady-state manifolds of strongly dissipative systems, where the {strong dissipation} adiabatically decouples the non-steady states from the dynamics, resulting in effective unitary dynamics~\cite{zanardi14}.

Then, it is a natural question whether the KCMs can be realized in quantum systems with engineered dissipation, instead of adding a strong interaction (energy penalty) to the Hamiltonian of isolated systems. In Ref.~\cite{stannigel14}, Stannigel \textit{et al.} proposed the simulation of the gauge theory by 
constraining the original Hamiltonian  within a certain subspace of the total Hilbert space, using the engineered classical noise and leveraging the Zeno effect. However, this study focused on the subspace satisfying the gauge condition and did not consider the entire decoherence-free subspaces of the system created by dissipation.
In addition, it was not discussed whether the effective dynamics obtained by noise (or dissipation) can be different from the one obtained solely by the Hamiltonian with strong interaction.

\begin{figure}
	\tikzstyle{circle1} = [circle,minimum width=1pt, minimum height=1pt, draw=black, fill=green!50!black,node distance=1.0cm]
	\tikzstyle{spring}=[decorate,decoration={zigzag,pre length=3pt,post
		length=3pt,segment length=4pt}]
	\tikzstyle{square} = [regular polygon,regular polygon sides=4, draw=black, fill=black!40,node distance=1pt]
	\tikzstyle{spring2}=[thick,decorate,decoration={aspect=0.6, segment length=4pt, amplitude=2pt,coil}]
	\label{diagram}
	\subfigure[]{
		\begin{tikzpicture}[scale=1.0]
			\centering
			\draw[fill=blue!60!white,draw=blue!70!black,fill opacity=0.3,  rounded corners=0.8pt=0.8pt] (0,0) rectangle (1,1);
			\draw node at (0.5,0.5) {$E_0$};
			\draw[fill=blue!60!white,draw=blue!70!black,fill opacity=0.3,  rounded corners=0.8pt] (1,0) rectangle (1.7,-0.7);
			\draw node at (1.4,-0.35) {$E_1$};
			\draw[fill=blue!60!white,draw=blue!70!black,fill opacity=0.3,   rounded corners=0.8pt=0.8pt] (1.7,-0.7) rectangle (2.2,-1.2);
			\draw node at (1.95,-0.95) {$E_2$};
			\filldraw[black] (2.4,-1.4) circle (1pt);
			\filldraw[black] (2.5,-1.5) circle (1pt);
			\filldraw[black] (2.6,-1.6) circle (1pt);
			\draw[fill=blue!60!white,draw=blue!70!black,fill opacity=0.3,   rounded corners=0.8pt=0.8pt] (2.7,-1.7) rectangle (3.1,-2.1);
			\draw[fill=blue!60!white,draw=blue!70!black,fill opacity=0.3,   rounded corners=0.8pt=0.8pt] (3.1,-2.1) rectangle (3.3,-2.3);
			\draw[thick, rounded corners=1pt] (-0.1,-2.4) rectangle (3.4,1.1);
			\node[circle1] (c1) at (0.5,2.5) {};
			\node[circle1] (c2) at (1.5,2.5) {};
			\node[circle1] (c3) at (2.5,2.5) {};
			\draw[spring] (-0.1,2.5) -- (c1);
			\draw[spring] (c1) node[above=2pt] {$i-1$} -- node[below=2pt] {$L_{i-1}$}(c2);
			\draw[spring] (c2) node[above=2.5pt] {$i$} node[above=23pt]{\scriptsize{$\dot{\rho}(t)=-i\left[ H+U\sum_i L_i,\rho(t)\right]$}}-- node[below=2pt] {$L_{i}$}(c3);
			\draw[spring] (c3) node[above=2pt] {$i+1$} -- (3.1,2.5);
			
		\end{tikzpicture}
		\label{hs_unitary}}
	\subfigure[]{
		\begin{tikzpicture}[scale=1.0]
			\centering
			\draw[fill=red!60!white,draw=red!70!black,fill opacity=0.3,   rounded corners=0.8pt=0.8pt] (0,0)  rectangle (1,1);
			\draw[fill=red!10!white,draw=red!10!black,fill opacity=0.3,   rounded corners=0.8pt=0.8pt] (1,0) rectangle (1.7,-0.7);
			\draw[fill=red!60!white,draw=red!70!black,fill opacity=0.3,   rounded corners=0.5pt=0.5pt] (1,0) rectangle (1.3,-0.3);
			\draw[fill=red!60!white,draw=red!70!black,fill opacity=0.3,   rounded corners=0.5pt=0.5pt] (1.3,-0.3) rectangle (1.5,-0.5);
			\draw[fill=red!60!white,draw=red!70!black,fill opacity=0.3,   rounded corners=0.5pt=0.5pt] (1.5,-0.5) rectangle (1.65,-0.65);
			\draw[fill=red!10!white,draw=red!10!black,fill opacity=0.3,    rounded corners=0.8pt=0.8pt=0.8pt] (1.7,-0.7) rectangle (2.2,-1.2);
			\draw[fill=red!60!white,draw=red!70!black,fill opacity=0.3,   rounded corners=0.5pt=0.5pt] (1.7,-0.7) rectangle (1.9,-0.9);
			\draw[fill=red!60!white,draw=red!70!black,fill opacity=0.3,   rounded corners=0.5pt=0.5pt] (1.9,-0.9) rectangle (2.05,-1.05);
			\draw[fill=red!60!white,draw=red!70!black,fill opacity=0.3,   rounded corners=0.5pt=0.5pt] (2.05,-1.05) rectangle (2.15,-1.15);
			\filldraw[black] (2.4,-1.4) circle (1pt);
			\filldraw[black] (2.5,-1.5) circle (1pt);
			\filldraw[black] (2.6,-1.6) circle (1pt);
			\draw[fill=red!60!white,draw=red!10!black,fill opacity=0.3,    rounded corners=0.8pt=0.8pt=0.8pt] (2.7,-1.7) rectangle (3.,-2.);
			\draw[fill=red!60!white,draw=red!10!black,fill opacity=0.3,   rounded corners=0.8pt=0.8pt] (3.,-2.) rectangle (3.2,-2.2);
			\draw[fill=red!60!white,draw=red!10!black,fill opacity=0.3,   rounded corners=0.8pt=0.8pt] (3.2,-2.2) rectangle (3.3,-2.3);
			\draw[thick, rounded corners=1pt] (-0.1,-2.4) rectangle (3.4,1.1);
			\draw node at (1.6,0.5) {$N_F=0$};
			\draw node at (2.3,-0.3) {$N_F=1$};
			\draw node at (2.75,-1.0) {$N_F=2$};
			\draw (0,3) -- (3.0,3);
			\node[circle1] (l1) at (0.5,3) {};
			\node[circle1] (l2) at (1.5,3) {};
			\node[circle1] (l3) at (2.5,3) {};
   			\node[square] (b0) at (0.0,2.2) {};
			\node[square] (b1) at (1.0,2.2) {};
			\node[square] (b2) at (2.0,2.2) {};
			\node[square] (b3) at (3.0,2.2) {};
                \draw[spring2] (l1) node[above=2pt] {$i-1$} -- (b0);
			\draw[spring2] (l1) -- (b1)node[below=4pt] {$L_{i-1}$};
                \draw[spring2] (l2) node[above=2pt] {$i$} -- node[below=0pt,left=2pt] {$\gamma$}(b1);
			\draw[spring2] (l2) node[above=15pt]{\scriptsize{$\dot{\rho}(t)=-i\left[ H,\rho(t)\right] + \mathcal{D}(\rho)$}} -- node[below=0pt,left=2pt] {$\gamma$}(b2)node[below=4pt] {$L_i$};
                \draw[spring2] (l3) node[above=2pt] {$i+1$} -- node[below=0pt,left=2pt] {$\gamma$}(b2);
			\draw[spring2] (l3) -- node[below=0pt,left=2pt] {$\gamma$}(b3)node[below=4pt] {$L_{i+1}$};
		\end{tikzpicture}
		\label{hs_dissipative}}
	\caption{Schematic illustration for the comparison between the Hamiltonian-constructed kinetically constrained models (KCMs) and Lindbladian-constructed KCMs. (a) Models constructed by the Hamiltonian $H' = H + U \sum_i L_i$ in the large interaction $U$ limit. We have disconnected energy subspaces, where $E_0$, $E_1$, $E_2,\cdots$ refer to the ground-state subspace, first excited-energy subspace, second excited-energy subspace of $H'$, and so on. (b) Models constructed by the Lindbladian $\mathcal{L}$ in Eq.~\eqref{lindblad} in the large dissipation limit. We have the block-diagonal structure of the effective Hamiltonian in the emergent  decoherence-free subspaces according to the number of frozen blocks $N_F$ present in the system. The Lindbladian-constructed KCMs have more separated structures than the 
 Hamiltonian-constructed KCMs due to the stronger dynamical constraints. We note that
the sectors within the energy subspaces of a Hamiltonian-constructed KCM may be further decomposed into multiple
subsectors, as in the case of the PXP model hosting scars.}
\end{figure}
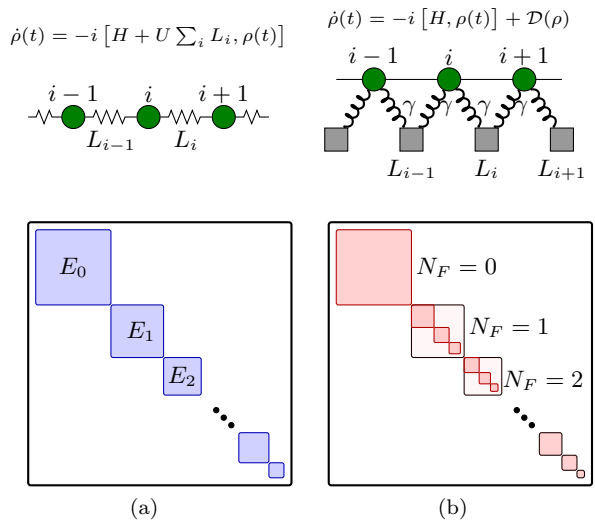

In this paper, we propose preparing KCMs using the engineered strong dissipation by considering the entire decoherence-free subspaces. 
{Focusing on the classical-noise engineering of the Gorini-Kossakowski-Sudarshan-Lindblad (GKSL) master equation with commuting Hermitian jump operators}, we discuss the emergent decoherence-free subspaces, which lead to the effective long-time unitary dynamics in the large dissipation limit. 
Importantly, we argue that the obtained dynamics (Lindbladian-constructed dynamics) can be KCMs that are more constrained than KCMs obtained from the Hamiltonian dynamics with strong interactions (Hamiltonian-constructed dynamics) defined from the same operators as the jump operators.
Figures~\ref{hs_unitary} and~\ref{hs_dissipative} schematically illustrate this comparison.

Specifically, we first consider a one-dimensional spin chain with two-site dissipation and introduce a ``PXQ model" as a Lindbladian-constructed KCM, which exhibits the free domain-wall motion with an additional  frozen-block structure.
This model is more constrained than the Hamiltonian-constructed counterpart, where such frozen blocks do not exist.
The PXQ model with a uniform magnetic field shows localization of the domain wall similar to the Wannier-Stark localization for a fermionic system under a tilted field, while the Hamiltonian-constructed counterpart shows a more complicated (de)localization behavior.
Moreover, coupling two PXQ chains with a uniform magnetic field and an inter-chain interaction, we discover a non-trivial partial delocalization of the domain wall for a certain parameter line, whereas the localization persists despite the interactions for the other parameter regimes.

We briefly mention some recent works that considered KCMs in dissipative quantum systems.
In Ref.~\cite{essler20}, it has been shown that the space of operators can be fragmented into exponentially many subspaces that are invariant under Lindbladian dissipative dynamics, and that each invariant is represented by an effective integrable Hamiltonian.
Recently, a method of embedding quantum many-body scars in the decoherence-free subspaces of the GKSL equation has been studied~\cite{wang23}. In this study, jump operators with local projectors are constructed so that the null space of the dissipators is spanned by the scar states only. Then, and the dissipative dynamics for generic initial states shows persistent coherent oscillations. In Ref.~\cite {li23}, the strong HSF of the bare Hamiltonian results in exponentially
many degenerate stationary states. A dephasing bath reduces the quantum fragmentation to a classical fragmentation with zero quantum correlations in the stationary state, while baths that preserve the structure of quantum fragmentation generate a highly entangled stationary state, i.e., preserved quantum fragmentation.
We stress that the aim of our manuscript is different from those results since we consider the emergence of the kinetic constraints for strong dissipation, which are absent for the original Hamiltonian alone.

The rest of the paper is organized as follows. We briefly overview the Markovian dynamics of open quantum systems represented in the extended Hilbert space in Sec.~\ref{sec_markovian}. In Sec.~\ref{sec_perturbation}, we provide a general formalism to derive an effective description of the dynamics in the emergent decoherence-free subspaces using the perturbation theory in the strong {dissipation} limit. We also compare the constrained dynamics in the Lindbladian-constructed models with their Hamiltonian-constructed counterparts. We then proceed to study a particular example, i.e., the PXQ model, and contrast it with the Hamiltonian-constructed counterpart in Sec.~\ref{sec_pxq}. Then, we present the results on localization and delocalization dynamics of two coupled PXQ chains in Sec.~\ref{sec_couple_pxq}.  Finally, we summarize our results and present concluding remarks in Sec.~\ref{sec_discussion}.  In appendix~\ref{app_pxp}, we present the Lindbladian construction of the PXP model.  We also perform the second-order perturbation theory to it in appendix~\ref{app_secondorder} and argue that the effective unitary description for Lindbladian-constructed models remains valid up to a timescale proportional to the {strength of the dissipation}.

\section{Markovian dynamics and its description in an extended Hilbert space}
\label{sec_markovian}
We focus on a quantum many-body system in which a Hamiltonian is locally coupled to a Markovian dissipation.
 The time evolution of the density matrix $\rho(t)$ of the system is described by the  GKSL master equation $\dot{\rho}(t) = \mathcal{L}(\rho(t))$, where the Liouvillian superoperator $\mathcal{L}(\rho(t))$ has the following form
\begin{eqnarray}
	\mathcal{L}(\rho(t))&=&-i\left[ H,\rho(t)\right] + \mathcal{D}(\rho)
	\label{lindblad}
\end{eqnarray}
with $\hbar$ set to unity. 
The dissipative term is given by
\begin{equation}
{\mathcal{D}(\rho)=\sum_{j=1}^{M}\gamma_j  \left( L_j \rho(t) L_j^{\dagger} - \frac{1}{2} \left\{L_j^{\dagger} L_j, \rho(t)\right\} \right),}
\end{equation}
where $L_j$ are the dissipation-inducing jump operators with a dissipation strength $\gamma_j$, and $M$ is the total number of jump operators. Here, $[A,B]$  ($\{A,B\}$) denotes the commutator (anti-commutator) between the operators $A$ and $B$.  {We consider uniform dissipation strength for all dissipation channels, i.e., $\gamma_j=\gamma$ for all $j$. We also assume that $H$, $L_j$, and $\gamma$ are time-independent.} 
The superoperator $\mathcal{L}(\rho)$ generates the completely-positive trace-preserving map $e^{\mathcal{L}t}$, with which the time evolution of the density matrix is described as $\rho(t)=e^{\mathcal{L}t}\rho(0)$~\cite{breuer_07, gardiner04}.

{While the GKSL equation often describes the dynamics of a system that is weakly coupled to an environment~\cite{breuer_07}, we instead focus on a situation where the GKSL equation is engineered by a noisy Hamiltonian.
Indeed, Ref.~\cite{chenu17} showed that the GKSL equation for the Hermitian jump operator $L_j$ is realized as the averaged dynamics of the noisy unitary dynamics, which is described as $H_\mathrm{noise}=H+\sum_j\sqrt{\frac{\gamma}{2}}\xi_j(t)L_j$. Here, the white classical noise $\xi_j(t)$ satisfies $\mathbb{E}_\mathrm{noise}[\xi_j(t)\xi_k(t')]=\delta_{jk}\delta(t-t')$. 
Throughout the manuscript, we consider such a setup with classical noise to realize our effective models under the strong dissipation.
}

It is convenient to rewrite the GKSL equation in a matrix representation, where the density matrix is represented as an element in an extended Hilbert space $\tilde{\mathcal{H}}=\mathcal{H}\otimes \mathcal{H}$:
\begin{eqnarray}
{ \dket{\rho}=\frac{1}{\mathcal{N}}\sum_{\sigma \tau} \rho_{\sigma \tau} \ket{\sigma} \otimes\ket{\tau}^{*},}
\end{eqnarray}
where $\mathcal{N}=\sqrt{\sum_{\sigma \tau} \lvert \rho_{\sigma \tau} \rvert ^2}$ denotes the normalization constant, {the state $\ket{\tau}^*$ is complex conjugation of the state $\ket{\tau}$}, and $\mathcal{H}$ represents the Hilbert space of the system.  For a $\mathcal{D}$-dimensional $\mathcal{H}$, the density matrix is simply mapped to a $\mathcal{D}^2$-dimensional vector $\dket{\rho}$  in the doubled Hilbert space $\mathcal{H}\otimes \mathcal{H}$. Then, the Lindblad equation takes the following equivalent form 
\begin{equation}
    \frac{d }{d t} \dket{\rho(t)} = \mathcal{L}\dket{\rho(t)} \equiv (\mathcal{L}_H + \mathcal{L}_D) \dket{\rho(t)}, 
    \label{eq_gksl_ex}
\end{equation}
with (see, e.g., \cite{yoshioka19})
\begin{eqnarray}\label{eq_lvec}
	 \mathcal{L}_H &=& -i\left(H\otimes I - I\otimes H^T\right),   \\ \mathcal{L}_D &=& \frac{\gamma}{2}\sum_{j=1}^{M}  \left( 2L_j \otimes L_j^{*}- L_j^{\dagger} L_j\otimes I - I \otimes L_j^{T} L_j^*\right),
\end{eqnarray}
where $\mathcal{L}_H$ and $\mathcal{L}_D$ are the unitary and dissipative parts of the Liouvillian $\mathcal{L}$ in the doubled Hilbert space.

Assuming that $\mathcal{L}$ is diagonalizable, the dynamics of the system can be described in terms of the spectrum of $\mathcal{L}$ as follows. The eigenvalue equations for right and left eigenvectors of $\mathcal{L}$ are given by
\begin{equation}
	\mathcal{L} \dket{\Lambda^R_\alpha}=\lambda_\alpha \dket{\Lambda_\alpha^R},~~\text{and}~~\mathcal{L}^{\dagger} \dket{\Lambda^L_\alpha}=\lambda_\alpha^* \dket{\Lambda_\alpha^L},
\end{equation}
where $\lambda_\alpha$ is the $\alpha$th eigenvalue with $\alpha=0,1,2, \cdots, \mathcal{D}^2-1$. The left and right eigenvectors corresponding to different eigenvalues are bi-orthogonal to each other; $\langle\langle \Lambda_\alpha^L\dket{\Lambda^R_\beta}=\delta_{\alpha,\beta}$. The time evolution of the density matrix can be written as
\begin{equation}
\dket{\rho(t)}=\sum_{\alpha=0}^{\mathcal{D}^2  -1} c_\alpha e^{\lambda_\alpha t}\dket{\Lambda_\alpha^R},
\end{equation}
where $c_\alpha=\langle\langle \Lambda_\alpha^L \dket{\rho_0}$ with  $\dket{\rho_0}$ being the vector for the initial density matrix.
It is important to note that the real part of each eigenvalue always satisfies $\textrm{Re}[\lambda_{\alpha}]\le 0$. Here, the eigenmodes are labeled in the descending order of the real part of eigenvalues as $0=\lambda_0\ge \textrm{Re}[\lambda_1] \ge \cdots \ge \textrm{Re}[\lambda_{\mathcal{D}^2-1}]$.  The stationary state $\dket{\rho_{\mathrm{ss}}}$ (satisfying $\mathcal{L}\dket{\rho_{\mathrm{ss}}}=0$) of the system is thus determined by the right eigenvector(s) of $\mathcal{L}$ corresponding to the zero eigenvalue(s). This is because all the other modes corresponding to eigenvalues having non-zero negative real part $\textrm{Re}[\lambda_{\alpha}] < 0$ eventually decay in the asymptotically long time.
Furthermore,  $\mathcal{L}$ has at least one eigenvalue that is equal to zero, $\lambda_0=0$~\cite{albert_14,nigro_2019}. This implies that there exists at least one stationary state. If $\lambda_{\alpha}=0$ has degeneracy of order $d$ (i.e., there exist $d$ independent right eigenvectors with zero eigenvalues), then there are $d$ independent stationary states towards which the system can evolve depending on the initial density matrix. This forms a $d$-dimensional stationary-state subspace $\tilde{\mathcal{H}}_\mathrm{ss}$, which is spanned by $\left\{\dket{\Lambda_{\alpha}^R}\right\}_{\alpha=0}^{d-1}$ with $\lambda_{\alpha}=0$.
Note that if there are non-zero eigenvalues whose real part is zero ($\lambda_\alpha\neq 0, \mathrm{Re}[\lambda_\alpha]=0$), the corresponding eigenvectors are regarded as {eternally oscillating modes without decay~\cite{buca19}}.

\section{Perturbation theory and emergent decoherence-free dynamics}
\label{sec_perturbation}
In this section, we consider a {strong dissipation limit, i.e., the dissipation strength $\gamma$} is much larger than the typical microscopic parameters $J_\mathrm{typ}$ of the Hamiltonian, $\gamma \gg J_\mathrm{typ}$,
and provide a general effective description of the dynamics in the emergent decoherence-free subspaces. 
In the large $\gamma$ limit, we shall use a perturbation theory  to derive an effective description. The Liouvillian in {Eq.~\eqref{eq_gksl_ex}} can be split into a large unperturbed part $\mathcal{L}_0$ and the unitary part as a perturbation $\mathcal{L}_1$, such that $\mathcal{L}=\mathcal{L}_0+\mathcal{L}_1$ with  $\mathcal{L}_1=\mathcal{L}_H$ and  $\mathcal{L}_0=\mathcal{L}_D$. 
The dynamics of the system for a time $t\gg 1/\gamma$ is governed by the eigenstates of $\mathcal{L}$ corresponding to the eigenvalues whose real parts satisfy $|\mathrm{Re}[\lambda_\alpha]|\ll \gamma$.

\subsection{Emergent decoherence-free subspaces}
For large $\gamma$, the dynamics is effectively constrained in the stationary-state subspace for $\mathcal{L}_0$ in the leading order of the perturbation.
We will call the corresponding subspace of the Hilbert space  the emergent decoherence-free subspace for $\mathcal{L}$, which becomes the exact decoherence-free subspace~\cite{lidar98,beige2000} for $J_\mathrm{typ}/\gamma \rightarrow 0$.

{In the present work, we will only consider jump operators that are Hermitian $L_j^{\dagger}=L_j$ and commutative with one another, $\left[L_i,L_j\right]=0$ for all $i$ and $j$, for the sake of simplicity. It turns out that we can easily characterize the emergent decoherence-free subspace in terms of the simultaneous eigenbasis of these jump operators.
Let us define a set of simultaneous eigenstates  for the jump operators as $\ket{f_1, f_2, \cdots f_M; m}\equiv\ket{\{f_j\};m}$, where $m$ is the label of the degeneracy.} The eigenvalue equation reads  
\begin{equation}\label{eq_ev_lind}
	L_j \ket{\{f_j\};m} = f_j \ket{\{f_j\};m},
\end{equation}
where $f_j$ is the eigenvalue for the jump operator $L_j$. Using the simultaneous eigenstates of the jump operators as a complete basis for $\mathcal{H}$, we can express a general density matrix as
\begin{equation}
{	\dket{\rho}=\frac{1}{\mathcal{N}}\sum_{\substack{\{f_j\}, m \\ \{g_j\},m'}} \rho_{f_j, g_j,m,m'} \ket{\{f_j\};m}\otimes\ket{\{g_j\};m'}^{*}.}
\end{equation}

Now, let us discuss the stationary–state subspace of $\mathcal{L}_0$ in terms of $\left\{\ket{\{f_j\};m}\right\}$. 
Each state $\dket{\rho^{(0)}_{\mathrm{ss}}}$ in the stationary–state subspace for $\mathcal{L}_0$ satisfies the following condition
\begin{equation}
{	\sum_{j=1}^{M}\left(2L_j \otimes L_j^{*}- L_j^{\dagger} L_j\otimes I - I \otimes L_j^{T} L_j^*\right)\dket{\rho^{(0)}_{\mathrm{ss}}}=0. }
\label{eq_ssmcond}
\end{equation}
{Moreover, the above condition for the stationary state reduces to
\begin{eqnarray}
	&&\sum_{\substack{\{f_j\}, \{g_j\} \\ m,m'}}\left(L_j \otimes I- I \otimes L_j^{*} \right)^2  \\\non
	&&~~~  \times(\rho_\mathrm{ss}^{(0)})_{f_j, g_j,m,m'}\ket{\{f_j\};m}\otimes\ket{\{g_j\};m'}^{*}
	=0. 
\end{eqnarray}
for all $j$. To obtain the above condition, we use the Hermiticity of the jump operator $L_j^\dagger = L_j$. Using Eq.~\eqref{eq_ev_lind} and $L_j^* \ket{\{g_j\};m}^* = g_j \ket{\{g_j\};m}^*$ in the above equation, one can obtain that $(\rho_\mathrm{ss}^{(0)})_{f_j, g_j,m,m'}$ is non-zero only when $f_j=g_j$ for all $j$.} This implies that the stationary-state subspace does not include states where the eigenstates with different eigenvalues $\{f_j\}$ of the Lindblad jump operators $\{L_j\}$ are mixed.
Thus, we can define a projection operator onto the stationary-state subspace as
\begin{equation}
{	\mathcal{P}= \sum_{\{f_j\}} p_{\{f_j\}} \otimes p_{\{f_j\}}^*,}
	\label{eq_ss_proj}
\end{equation}
where $p_{\{f_j\}}$ is the projection operator to the  degenerate eigensubspace of $L_j$ with eigenvalues $\{f_j\}$,
\begin{equation}\label{eq_proj_des}
	p_{\{f_j\}}=\sum_m \ket{\{f_j\};m}\bra{\{f_j\};m},
\end{equation}
{and $p_{\{f_j\}}^*$ is the complex conjugation of it.}

\subsection{Leading-order term: effective Hamiltonian}
\label{sec_firstorder}

The eigenspace for zero eigenvalues of $\mathcal{L}_0$ may, in general, have a large degeneracy. Thus, the dimension of the stationary-state subspace defined by the projection operator $\mathcal{P}$ in Eq.~\eqref{eq_ss_proj} can be large.  In the large $\gamma$ limit, the eigenspectrum of $\mathcal{L}_0$ has a large spectral gap ($\Delta\sim\gamma$) between zero eigenvalues ($\lambda_{\alpha}=0$) and non-zero eigenvalues ($\lambda_{\alpha}\neq 0$). The effect of the perturbation $\mathcal{L}_1$ term lifts the degeneracies through transition processes within the stationary-state subspace of $\mathcal{L}_0$. We perform the Schrieffer–Wolff version of degenerate perturbation theory~\cite{bravyi11,kessler12_b} to derive an effective Liouvillian $\mathcal{L}_{\textrm{eff}}$ that describes the dynamics of the system in the subspace of $\mathcal{L}$ with $\mathrm{Re}[\lambda_\alpha]\simeq 0$. The first-order term in the perturbation theory provides the following effective Liouvillian,
\begin{eqnarray}
	\mathcal{L}_{\mathrm{eff}}^{\left(1\right)}=\mathcal{P}\mathcal{L}_1\mathcal{P}=-i\mathcal{P}\left(H\otimes I - I\otimes H^T\right)\mathcal{P}.
\end{eqnarray}
Using Eq.~\eqref{eq_ss_proj} in the above equation, we get
{\begin{eqnarray}	\mathcal{L}_{\mathrm{eff}}^{\left(1\right)}&=&
	-i\sum_{\{f_j\}} \big( p_{\{f_j\}}Hp_{\{f_j\}}\otimes p_{\{f_j\}}^* \non \\ 
	&&~~~~~ ~~~ - p_{\{f_j\}}\otimes p_{\{f_j\}}^*H^Tp_{\{f_j\}}^*\big), 
\end{eqnarray}}
where we have used $p_{\{f_j\}}p_{\{g_j\}}=\delta_{f_j,g_j}p_{\{f_j\}}$. The right-hand side of the above equation implies that the Liouvillian essentially describes a unitary dynamics in the subspace characterized by every $\{f_j\}$, which is regarded as the emergent decoherence-free subspace for $\mathcal{L}$. In other words, if we consider an initial state in one of the projected Hilbert spaces obtained by the projection operator $p_{\{f_j\}}$, the eventual dynamics  can be described by an effective Hamiltonian
\begin{equation}
	H_{\mathrm{eff}}^{\mathrm{lind}}=p_{\{f_j\}}Hp_{\{f_j\}}
	\label{eq_effh_ss}
\end{equation}
for a certain timescale when the first-order perturbation theory is valid.
This is a formal expression of the Lindbladian-constructed effective Hamiltonian.
As shown in the following sections, $H_\mathrm{eff}^{\mathrm{lind}}$ can naturally be a KCM for appropriate dissipation $\{L_j\}$.
We note that the next order of the perturbation can systematically be evaluated, as demonstrated in Appendix~\ref{app_secondorder} for a certain model.


\subsection{Comparison with the unitary case}
\label{sec_comparison}
Before establishing some examples of the Lindbladian-constructed dynamics in Eq.~\eqref{eq_effh_ss}, we discuss the distinction  from the conventional method of obtaining effective dynamics from the time-independent Hamiltonian alone.
We specifically consider the following form of the Hamiltonian as a naive counterpart of the Lindbladian construction:
\begin{equation}
	H'= H + U \sum_{j=1}^M L_j,
\end{equation}
where $H$ is the same Hamiltonian generating unitary dynamics in the GKSL equation and $L_j$ are the same Hermitian operators that we have considered as jump operators. In the large $U$ limit, we can treat $H$ as a perturbation.  The simultaneous eigenstates of $\{L_j\}$ given in Eq.~\eqref{eq_ev_lind} form different degenerate energy subspaces separated by a large gap ($\Delta\sim U$). One can obtain an effective Hamiltonian in the ground-state energy subspace via the Schrieffer-Wolff transformation  as
\begin{equation}
{H}_\mathrm{eff}^{\mathrm{ham}}=p_0Hp_0+E_0,
\label{unitary_eff_ham}
\end{equation}
where $p_0$ is the projection to the ground-state energy subspace with a (constant) energy $E_0$ in Fig.~\ref{hs_unitary}.
Likewise, if the initial state is in an excited-state energy subspace with energy $E_{\sum_jf_j}=U\sum_jf_j$, the effective dynamics is given by 
\begin{equation}\label{excited_eff}
{H}_\mathrm{eff}^{\mathrm{ham}}=p_{\sum_jf_j}Hp_{\sum_jf_j}+E_{\sum_jf_j},
\end{equation}
where $p_{\sum_jf_j}$ is the projection operator onto the excited-energy subspace with the (constant) energy $E_{\sum_jf_j}$.
These effective dynamics predict the actual dynamics for timescales that increase with increasing $U$, as rigorously discussed in Refs.~\cite{Gong20,Gong20_pra,Gong22}.

We compare the above Hamiltonian-constructed effective dynamics with the Lindbladian-constructed effective dynamics in Eq.~\eqref{eq_effh_ss}.
First, the ground-state subspace in Eq.~\eqref{unitary_eff_ham} corresponds to $\{f_j^\mathrm{min}\}$ where $f_j^\mathrm{min}$ is the minimum eigenvalue of $L_j$ in Eq.~\eqref{eq_ev_lind} for every $j$. 
In such a way, $f_j^\mathrm{min}$ is uniquely determined for each $j$, and we find $p_{\{f_j^\mathrm{min}\}}=p_0$.
In other words, the Hamiltonian-constructed effective dynamics \eqref{unitary_eff_ham} in the ground-state energy subspace is equivalent to the Lindbladian-constructed effective dynamics \eqref{eq_effh_ss} for the subspace satisfying $\{f_j\}=\{f_j^\mathrm{min}\}$.

In contrast, the Lindbladian-constructed effective dynamics with $\{f_j\}\neq\{f_j^\mathrm{min}\}$ is in general distinct from the Hamiltonian-constructed effective dynamics \eqref{excited_eff} in the excited-state energy subspace. Indeed, while $p_{\{f_j\}}$ determines the constraint for every $j$, $p_{\sum_jf_j}$ only leads to a constraint for the sum of $f_j$. 
Therefore, the Lindbladian-constructed effective dynamics is, in general, more strongly constrained than the Hamiltonian-constructed one, as exemplified in the next section.

\subsection{Frozen blocks}\label{frozen}
Before ending this section, we briefly mention  the frozen blocks and the block-diagonal structure of the Hamiltonian for the Lindbladian-constructed effective dynamics.
For this purpose, we assume that each jump operator $L_j$ acts on a local region $X_j$ for a spin system, where the Hilbert space is given by a tensor product of local Hilbert spaces for the spin.
Then, each $f_j$ is obtained from the diagonalization of $L_j$ within $X_j$ as 
\begin{align}
L_j\ket{f_j;s_j}_{X_j}=f_j\ket{f_j;s_j}_{X_j}\:(1\leq s_j\leq S_{j,f_j}),
\end{align}
{where $\ket{f_j;s_j}_{X_j}$ is in $\mathcal{H}_{X_j}$, the local Hilbert space for $X_j$. The quantity $S_{j, f_j}$ is the number of degeneracies (i.e., the maximum value of the degeneracy label $s_j$) for the eigenstate $\ket{f_j; s_j}_{X_j}$.}
In particular, if $S_{j,f_j}=1$ (no degeneracy), the region $X_j$ is only spanned by {$\ket{f_j}_{X_j}\otimes \ket{f_j}_{X_j}^*$} in the corresponding  emergent decoherence-free subspace.
This means that this sub-region corresponds to a frozen block where no dynamics takes place.

Then, in this case, the effective Hamiltonian in the emergent decoherence-free subspaces is block-diagonalized by the different number ($N_F$) of frozen blocks. Furthermore, the position of the frozen blocks can further split a particular $N_F$-sector into many disconnected subsectors, as schematically represented in Fig.~\ref{hs_dissipative}.

We note that when $S_{j,f_j^\mathrm{min}}>1$ for every $j$, there are no frozen blocks in the subspace projected by $p_{\{f_j^\mathrm{min}\}}$.
Therefore, the Hamiltonian-constructed effective dynamics \eqref{unitary_eff_ham} in the ground-state subspace corresponds to the no-frozen-block subspace for the Lindbladian-constructed model in this case.

\section{Example: PXQ Model}
\label{sec_pxq}
Let us now consider a particular example by considering the Hamiltonian of a simple spin-1/2 non-interacting system,
\begin{eqnarray}
	H=\sum_{i=2}^{N-1} \left(J \sigma_i^x-  \frac{h}{2} \sigma_i^z\right) ,
	\label{eq_ham}
\end{eqnarray}
{where $\sigma_i^{x,y,z}$ are the Pauli's spin operators at the $i$th site. We consider the lattice spacing to be unity throughout this manuscript.} The parameter $h$ is the longitudinal field, $J$ is the transverse field, and $N$ is the total number of spins. We omit the field for the first and $N$th spins in the chain. 
Consequently, these boundary spins after time evolution remain unchanged from those for the initial state, as discussed in the subsequent sections of the paper.

We first consider  Markovian jump operators given by $L_i=Q_i P_{i+1}$ with open boundary condition (OBC) $i=1,2,\cdots N-1$, where $Q_i=(1 + \sigma^z_i)/2$ and $P_i=1-Q_{i}$. The operator $Q_i$ ($P_i$) acts on the $i$th spin as $Q_i\ket{1_i}=\ket{1_i}$ ($P_i \ket{1_i}=0$) and $Q_i \ket{0_i}=0$ ($P_i \ket{0_i}=\ket{0_i}$). Here, we represent an up (down) spin state by $\ket{1}$ ($\ket{0}$). 
As mentioned in Sec.~\ref{sec_markovian}, this kind of Markovian jump operator can be realized in quantum many-body systems using an appropriate classical white noise~\cite{chenu17}. 

To obtain the emergent decoherence-free subspaces, we first calculate the eigenspectrum of the jump operators $L_i$. 
Considering a simple computational  basis $\ket{n_1, n_2,\cdots, n_L}$, where $n_i$ is either $0$ or $1$, we find that they become the eigenstates for $L_i$ as
$
L_i \ket{n_1, n_2,\cdots n_L}=f_i\ket{n_1, n_2,\cdots, n_L}.
$
Moreover, since $L_i$ acts on the local region $X_i=\{i,i+1\}$, we can focus on the two sites $\ket{n_in_{i+1}}_{X_i}$.
Then, we have two degenerate eigenspaces for each jump operator $L_i$ with eigenvalue $f_i=0$ and $f_i=1$.   Indeed, the eigenstates of $L_i$ are given by $\ket{00}, \ket{01},\ket{11}$ and $\ket{10}$ (the subscript $X_i$ is omitted for brevity). 
The first three eigenstates are degenerate with eigenvalue $f_i=0$ (with $S_{i,0}=3$), and the eigenvalue corresponding to the eigenstate $\ket{10}$ is $f_i=1$ (with $S_{i,1}=1$). Thus, the stationary-state subspace for $\mathcal{L}_0$ (which leads to the emergent decoherence-free subspaces for $\mathcal{L}$) is spanned by the  states whose local configurations are given in the following:
\begin{eqnarray}\label{eq_edfs_pxq}
	\Big\{\ket{00}\otimes\ket{00}, \ket{00}\otimes\ket{01}, \ket{00}\otimes\ket{11},  \ket{01}\otimes\ket{00}, \non\\  \ket{01}\otimes\ket{01}, \ket{01}\otimes\ket{11},  \ket{11}\otimes\ket{00}, \ket{11}\otimes\ket{01}, \non\\ \ket{11}\otimes\ket{11}, ~\text{and}~\ket{10}\otimes\ket{10}\Big\}. 
\end{eqnarray}

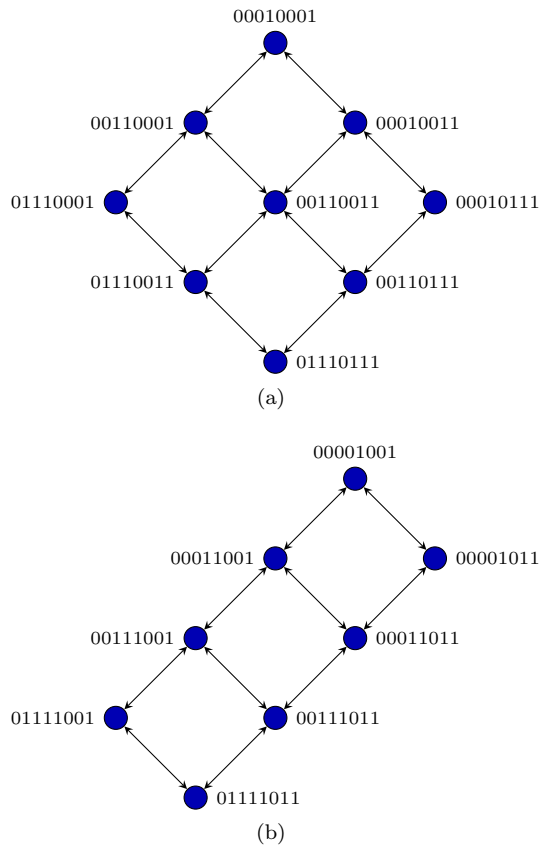
\begin{figure}
	\centering 
	\tikzstyle{circle1} = [circle, rounded corners,text centered, draw=black, fill=blue!70!black,node distance=1.5cm]
	\tikzstyle{arrow} = [ <->,>=stealth]
	\subfigure[]{\begin{tikzpicture}   
			\node[circle1, label=above:\scriptsize{{$00010001$}}](a0){\scriptsize{}};
			\node[circle1, below left of=a0,label=left:\scriptsize{{$00110001$}}](a1){};
			\node[circle1,below right of=a0,label=right:\scriptsize{{$00010011$}}](a2){};
			\node[circle1,below left of=a1,label=left:\scriptsize{{$01110001$}}](a3){};
			\node[circle1,below left of=a2,label=right:\scriptsize{{$00110011$}}](a4){};
			\node[circle1,below right of=a2,label=right:\scriptsize{{$00010111$}}](a5){};
			\node[circle1,below right of=a3,label=left:\scriptsize{{$01110011$}}](a6){};	
			\node[circle1,below left of=a5,label=right:\scriptsize{{$00110111$}}](a7){};		
			\node[circle1,below right of=a6,label=right:\scriptsize{{$01110111$}}](a8){};		
			
			\draw[arrow] (a0) -- (a1);
			\draw[arrow] (a0) -- (a2);
			\draw[arrow] (a1) -- (a3);
			\draw[arrow] (a1) -- (a4);
			\draw[arrow] (a2) -- (a4);
			\draw[arrow] (a2) -- (a5);
			\draw[arrow] (a3) -- (a6);
			\draw[arrow] (a4) -- (a6);
			\draw[arrow] (a4) -- (a7);
			\draw[arrow] (a5) -- (a7);
			\draw[arrow] (a6) -- (a8);
			\draw[arrow] (a7) -- (a8);
			
	\end{tikzpicture}}
	
	\subfigure[]{\begin{tikzpicture}   
			\node[circle1, label=above:\scriptsize{{$00001001$}}](a0){\scriptsize{}};
			\node[circle1, below left of=a0,label=left:\scriptsize{{$00011001$}}](a1){};
			\node[circle1,below right of=a0,label=right:\scriptsize{{$00001011$}}](a2){};
			\node[circle1,below left of=a1,label=left:\scriptsize{{$00111001$}}](a3){};
			\node[circle1,below right of=a1,label=right:\scriptsize{{$00011011$}}](a4){};
			\node[circle1,below left of=a3,label=left:\scriptsize{{$01111001$}}](a5){};
			\node[circle1,below right of=a3,label=right:\scriptsize{{$00111011$}}](a6){};	
			\node[circle1,below right of=a5,label=right:\scriptsize{{$01111011$}}](a7){};		
			
			\draw[arrow] (a0) -- (a1);
			\draw[arrow] (a0) -- (a2);
			\draw[arrow] (a1) -- (a3);
			\draw[arrow] (a1) -- (a4);
			\draw[arrow] (a2) -- (a4);
			\draw[arrow] (a3) -- (a5);
			\draw[arrow] (a4) -- (a6);
			\draw[arrow] (a3) -- (a6);
			\draw[arrow] (a6) -- (a7);
			\draw[arrow] (a5) -- (a7);
			
	\end{tikzpicture}}
	\caption{Graph representation for the dynamics of the PXQ model (with open boundary condition) of system size $N=8$ for two different initial states: (a) $\ket{00010001}$ and (b) $\ket{00001001}$. 
 Here, we have fixed the first and $N$th boundary spins to $0$ and $1$, respectively. Both of the initial states lie in the subspace containing only one frozen block $N_F = 1$. However, the position of the frozen block is different in the two initial states, which results in two different (disconnected) graphs. These graphs correspond to two disconnected subsectors within the sector $N_F = 1$ of the emergent decoherence-free subspaces.}
	\label{fig_ghrap}
\end{figure}

\subsection{PXQ model without a longitudinal field \texorpdfstring{($h=0$)}{h=0}}
Let us begin with the case when the unitary part of the GKSL equation is governed by the single-term Hamiltonian $H=\sum_{i=2}^{N-1} J \sigma_i^x$ with $h=0$. 
In this case, the associated perturbation term in the Liouvillian,
\begin{eqnarray}
    \mathcal{L}_1&=&-iJ\sum_{i=2}^{N-1}\left(\sigma_i^x\otimes I - I\otimes \sigma_i^x\right),
    \label{perturbation_pxq}
\end{eqnarray}
will generate transitions among the stationary states of $\mathcal{L}_0$. 
Looking at the allowed configurations for two adjacent spins in Eq.~\eqref{eq_edfs_pxq}, the only possible transitions generated by $\mathcal{L}_1$ within the stationary-state subspace for $\mathcal{L}_0$  are the following:
\begin{eqnarray}
	\ket{001}\otimes\ket{\cdot \cdot \cdot} \longleftrightarrow \ket{011}\otimes\ket{\cdot \cdot \cdot }, \\
		\ket{\cdot \cdot \cdot}\otimes\ket{001} \longleftrightarrow \ket{\cdot \cdot \cdot}\otimes\ket{011}.
\end{eqnarray}
Restricting to the original Hilbert space, the only allowed transition is $\ket{001}\longleftrightarrow\ket{011}$. Thus, considering the matrix elements for the transitions, one can write an effective Hamiltonian describing the dynamics of the system in the emergent decoherence-free subspaces in the  following form
\begin{equation}
	H_{\mathrm{eff}}=J\sum_{i=2}^{N-1} P_{i-1}\sigma_i^x Q_{i+1}.
 \label{ham_pxq}
\end{equation}
We call the above Hamiltonian the PXQ model. Here, we  mention that the above Hamiltonian $H_{\mathrm{eff}}$ (having local interactions) is different from $H_{\mathrm{eff}}^{\mathrm{lind}}$ defined in Eq.~\eqref{eq_effh_ss}, which generally has nonlocal interactions. However, we have
\begin{eqnarray}
    H_{\mathrm{eff}}^{\mathrm{lind}}=p_{\{f_i\}}Hp_{\{f_i\}}=p_{\{f_i\}}H_{\mathrm{eff}}p_{\{f_i\}},
\end{eqnarray}
where $p_{\{f_i\}}$ is defined in Eq.~\eqref{eq_proj_des}.
Therefore, if we focus on a single emergent decoherence-free subspace spanned by {$p_{\{f_i\}}\otimes p_{\{f_i\}}^*$}, the dynamics caused by $H_\mathrm{eff}^\mathrm{lind}$ and $H_\mathrm{eff}$ are essentially equivalent. {We note that this kind of directional KCMs is also observed via purely coherent processes in a Rydberg atomic array with a staggered configuration of atomic position and classical drive fields~\cite{valenciatortora23}.}

As an example of the time evolution in this model, we present the dynamics of the system with $N=8$ spins from the initial state $\ket{00000001}$, where we have respectively fixed the first and $N$th spins to $0$ and $1$:
\begin{eqnarray}
		\ket{00000001} \leftrightarrow \ket{00000011}\leftrightarrow \cdots \leftrightarrow\ket{01111111}. 
\label{dynamics_f0}
	\end{eqnarray}
One can realize that the constrained dynamics shown in the above equation for the initial state $\ket{00000001}$ are essentially the dynamics of a free particle if we look at the movement of the domain wall $``01"$. However, the dynamics in the entire emergent decoherence-free subspaces is not the simple free particle dynamics because of the existence of frozen blocks, as discussed below.

Let us consider the structure of the constrained dynamics generated by the above effective Hamiltonian in detail. We see that the configuration $\ket{\cdots 10 \cdots}\otimes\ket{\cdots 10 \cdots}$ for two adjacent spins is allowed in the emergent decoherence-free subspaces.
Nonetheless, the Hamiltonian in Eq.~\eqref{ham_pxq} cannot change this configuration. Thus, the configuration $\ket{\cdots 10 \cdots}\otimes\ket{\cdots 10 \cdots}$ is a frozen block (recall that $S_{i,1}=1$, which is the condition for the frozen block).
Therefore, the emergent decoherence-free subspaces will split into several disconnected sectors according to the number of frozen blocks. Furthermore, two regions separated by a frozen block cannot talk to each other.  Each sector determined by the number of frozen blocks will also be decomposed into many disconnected subsectors depending upon the position of the frozen blocks. 

In the dynamics shown in Eq.~\eqref{dynamics_f0}, no frozen block is present in the initial state. Thus, the sector $N_F=0$ has no disconnected subsectors, and the dynamics in this sector is mapped to the simple free-particle dynamics. However, if we consider a sector with a higher number of frozen blocks, it will split into many disconnected subsectors. For example, let us consider a frozen-block configuration ``$10$" for two consecutive spins at some particular position on the chain. Then, the system follows a dynamics in which two segments of the chain on two sides of the frozen block do not interfere. In this case, the sector with $N_F=1$ can further be decomposed into different subsectors depending upon the position of the frozen block. This is shown in Fig.~\ref{fig_ghrap} in a graph representation.	However, all of the subsectors are integrable.

\subsection{Comparison with the Hamiltonian construction: PXQ–QXP model}
We emphasize that one cannot obtain the constrained dynamics discussed above from the naive time-independent Hamiltonian approach, i.e., in the unitary dynamics by the Hamiltonian  $H'=H+U\sum_{i=1}^{N-1} Q_i P_{i+1}$. For $H=\sum_{i=2}^{N-1} J \sigma_i^x$ with $h=0$, the Hamiltonian $H'$ is essentially the quantum Ising chain,
\begin{equation}\label{eq_qic}
    H'= - \frac{U}{4} \sum_{i=1}^{N-1}\sigma^z_i \sigma^z_{i+1} +J\sum_{i=2}^{N-1}\sigma_i^x + \frac{U}{4}\left(\sigma^z_1 - \sigma^z_N\right),
\end{equation}
where we omit the constant energy term $(N-1)U/4$. In the strong Ising coupling limit, $U \gg J$, the unperturbed part $H_0=-U/4 \sum_{i=1}^{N-1}\sigma^z_i \sigma^z_{i+1}$ possesses highly degenerate energy sectors labeled by the number of domain walls present in the spin configurations. These degenerate energy sectors are separated by a large gap.
Thus, transitions between different degenerate energy sectors by the perturbation term $H_1 = J\sum_{i=2}^{N-1}\sigma_i^x$ are suppressed. It results in effective dynamics describing transitions within each of the energy subspaces. Importantly, unlike the Lindbladian-constructed PXQ model, there are no local frozen blocks in the dynamics of this model.

For large enough $U$, the configuration ``$10$" for two adjacent spins is energetically prohibited in the ground-state manifold. The ground–state energy subspace consists of the states $\{\ket{00}, \ket{01}, \ket{11}\}$. 
This is the degenerate energy subspace corresponding to the single domain-wall case of $H_0$ for the quantum Ising chain~\eqref{eq_qic}, when we fix the first and $N$the boundary spins as 0 and 1. At the first order in perturbation theory, the dynamics in the ground-state energy subspace is then governed by the effective Hamiltonian 
\begin{equation}
	H_{\mathrm{eff}}=J\sum_{i=2}^{N-1} P_{i-1}\sigma_i^x Q_{i+1},
 \label{gs_pxq}
\end{equation}
with OBC. This is exactly the same as the no-frozen block sector $N_F=0$ of the effective Hamiltonian in the Lindbladian-constructed dynamics. 
This is also understood from $f_j^\mathrm{min}=0$ with $S_{j,f_j^\mathrm{min}}=3>1$, as discussed in Sec.~\ref{frozen}.

The main difference in the dynamics between the Hamiltonian-constructed model and the Lindbladian-constructed model is prominent if we look at the excited-energy subspace of the effective dynamics.  For example, if we consider an initial state with one ``$10$" excitation present in the configuration and fix the first and $N$th spins to $0$, the Hamiltonian-constructed effective dynamics show significantly different behavior from the Lindbladian-constructed one, as the following transitions are energetically allowed:
\begin{eqnarray}
\ket{00010000}\leftrightarrow\ket{00011000}\leftrightarrow\cdots\leftrightarrow\ket{00111100}\leftrightarrow\cdots, \non \\
\end{eqnarray} 
and so on~\footnote{Note that the Lindbladian-constructed dynamics leads to $\ket{00010000}\leftrightarrow\ket{00110000}\leftrightarrow\ket{01110000}$ from this initial state and the boundary condition}. This refers to the effective dynamics in the first excited-state energy subspace of the Hamiltonian case. Using standard perturbation theory, an effective Hamiltonian can be obtained to describe the dynamics in the higher-energy subspaces, as shown in Refs.~\cite{mazza19,lerose20}. The effective Hamiltonian for the Hamiltonian construction is given by
\begin{equation}
	H_{\mathrm{eff}}=J\sum_{i=2}^{N-1} \left(P_{i-1}\sigma_i^x Q_{i+1} + Q_{i-1}\sigma_i^x P_{i+1}\right),
 \label{gs_pxq_qxp}
\end{equation}
which we call the PXQ–QXP model in this manuscript to contrast with the PXQ model. Note that the PXQ-QXP model can also describe the dynamics in the ground-state subspace, which is described by Eq.~\eqref{gs_pxq}, because the second term vanishes for this subspace.
The difference between the Hamiltonian constructed PXQ–QXP model and the Lindbladian constructed PXQ model will be more prominent when we apply an additional uniform longitudinal field ($h\neq 0$), as we will discuss in the next section. 

\begin{figure}
\centering
\subfigure[]{%
	\includegraphics[width=.23\textwidth,height=4cm]{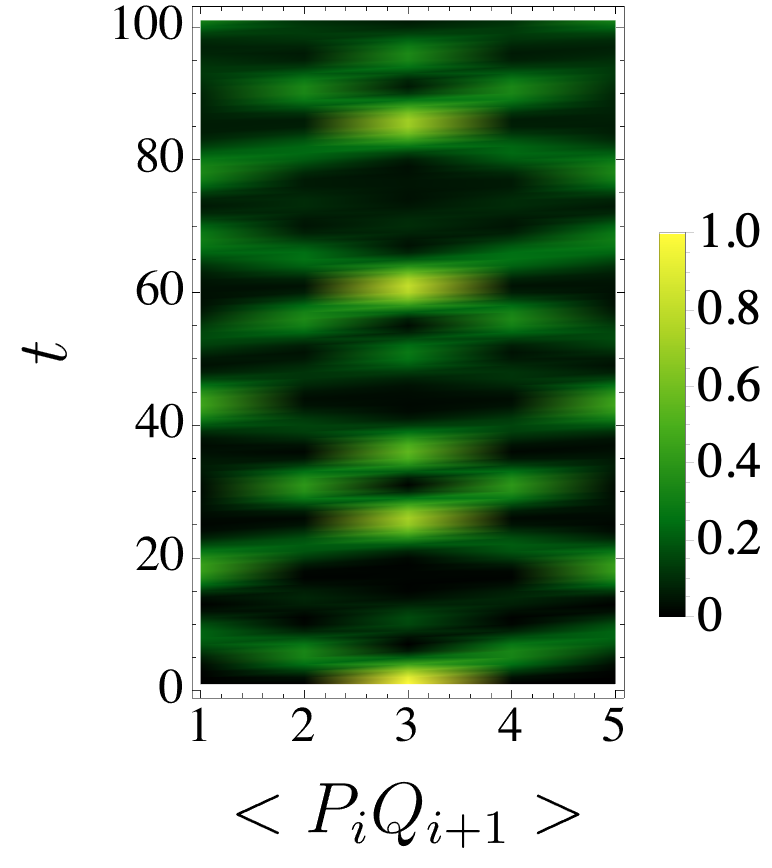}
	\label{fig_stark_h0}}
\subfigure[]{%
	\includegraphics[width=.23\textwidth,height=4cm]{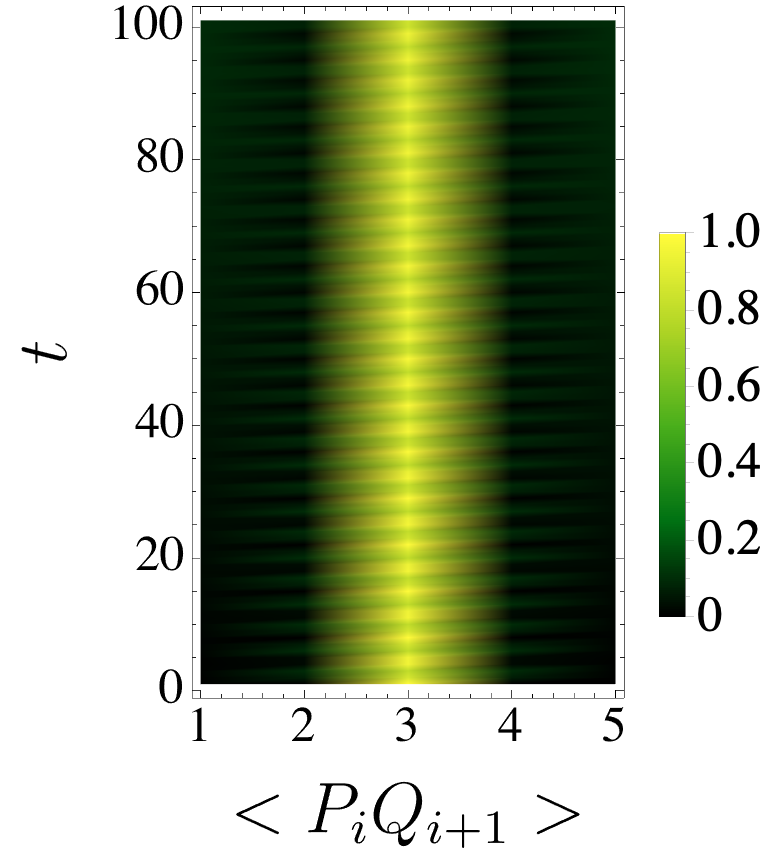}
	\label{fig_stark_h10}}
\caption{ (Color online) Time evolution of the ``$01$" domain-wall position  represented by the quantity $\langle P_iQ_{i+1}\rangle$ for all sites of the chain. We choose the initial configuration as $\dket{\rho_0}=\ket{000111}\otimes\ket{000111}$. (a) The domain wall delocalizes over the system when the longitudinal field value is zero, $h=0$. (b) Localization of the domain wall occurs for a non-zero value of the longitudinal field $h=3$. In both cases, we consider that  the total system size is $N=6$, {hopping strength $J=1$}, and the {dissipation strength} $\gamma=10^3$. }
\label{fig_stark}
\end{figure}

\subsection{Domain-wall localization in  the PXQ model with a longitudinal field \texorpdfstring{($h\neq0$)}{}}
\label{stark_localization}
Here, we study the localization of a domain wall,  which can occur in the PXQ model in the presence of a uniform longitudinal field $h$ in the Hamiltonian $H$ [Eq.~\eqref{eq_ham}]. Figure~\ref{fig_stark} shows the time evolution of the domain wall by calculating $\langle P_i Q_{i+1} \rangle$ for all the lattice sites by solving the GKSL master equation [Eq.~\eqref{eq_gksl_ex}] with Hamiltonian $H$ and $L_i=Q_i P_{i+1}$. The quantity $\langle P_i Q_{i+1} \rangle$ takes the value $1$ only for a ``$01$" domain wall, i.e., when the domain wall is formed by a down spin at left and an up spin at right. We choose an initial state $\dket{\rho_0}=\ket{000111}\otimes\ket{000111}$ with zero frozen block. We see that in the absence of a longitudinal field, the domain wall delocalizes in the system.  On the other hand, the domain wall remains localized for a non-zero value of the longitudinal field. 
We stress that this localization occurs for a uniform Hamiltonian and requires strong dissipation.

\begin{figure}
\centering
\includegraphics[width=.4\textwidth,height=.18\textwidth]{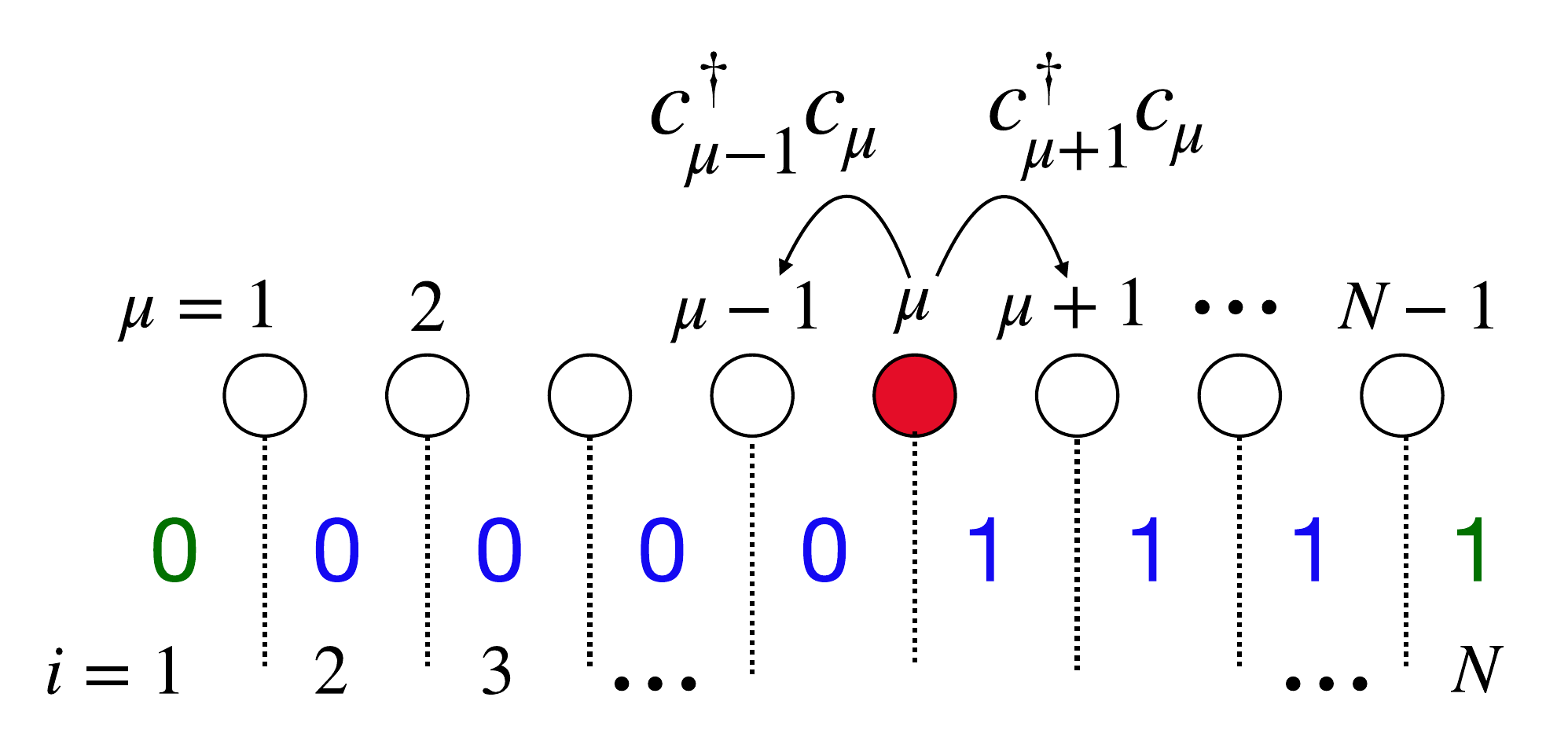}
\caption{ Schematic illustration for the quasi-particle representation of the domain-wall dynamics in the actual spin configuration in the PXQ model. The indices $i$ represent the sites for the original spins, and $\mu$ represents the position of the quasi-particle (position of bonds in the original spin chain). Here, we consider $N_F=0$ sector and the open boundary condition by fixing first and $N$th spins (shown in green) to $0$ and $1$, respectively.
}
\label{fig_diagram_tb}
\end{figure}

The dynamics of the domain wall presented in Fig.~\ref{fig_stark} can be understood through the mechanism of the single-particle Wannier-Stark localization~\cite{wannier62}.
For $h=0$, the dynamics in the zero–frozen-block sector of the emergent decoherence-free subspaces can be thought of as the free movement of the ``$01$" domain wall through the chain. In the spirit of an effective quasi-particle description~\cite{lin17}, the dynamics of the single domain wall can be described by the free quasi-particle hopping. In terms of fermionic operators $c_\mu$ and $c_\mu^{\dagger}$, the equivalent effective Hamiltonian is then given by $\tilde{H}=J\sum_{\mu=1}^{N-2}  (c_{\mu}^{\dagger} c_{\mu+1} + \mathrm{h.c.})$ with OBC. The index $\mu$ represents the positions of the quasi-particle excitation (the position of the bonds in the original spin chain).  This is schematically illustrated in Fig.~\ref{fig_diagram_tb}.
More interestingly,  if we apply a uniform longitudinal  field $h\neq 0$, the equivalent Hamiltonian for  the quasi-particle takes the following form
\begin{equation}
\tilde{H}=\sum_{\mu=1}^{N-2} J\left( c_\mu^{\dagger} c_{\mu+1} + \mathrm{h.c.}\right) + \sum_{\mu=1}^{N-1}h \mu c_\mu^{\dagger} c_\mu,
\end{equation}
where we neglect the constant energy term $-Nh/2$. The uniform magnetic field for the original spin system serves as a linear potential for the fermionic model. For the above Hamiltonian, each single-particle eigenstate, $c_\mu^{\dagger}\ket{0}$, is localized around some site with an inverse localization length given by $\xi ^{-1} = 2\sinh ^{-1} (h/2)$~\cite{Nieuwenburg19}. In sectors with a higher number of frozen blocks in the emergent decoherence-free subspaces, this phenomenon of single-particle Wannier-Stark localization (localization of domain walls in the spin language) will occur in each of the subblocks separated by the frozen blocks.

\begin{figure*}
\centering

\subfigure[]{%
	\includegraphics[width=.4\textwidth,height=.35\textwidth]{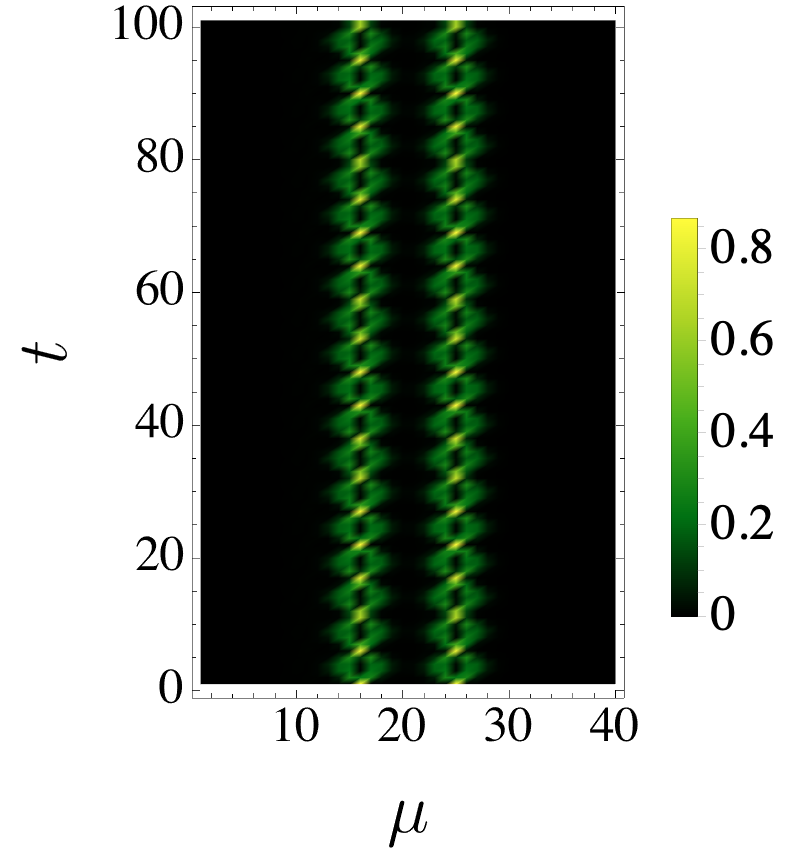}
	\label{fig_dw_separate}}
 \subfigure[]{%
	\includegraphics[width=.4\textwidth,height=.35\textwidth]{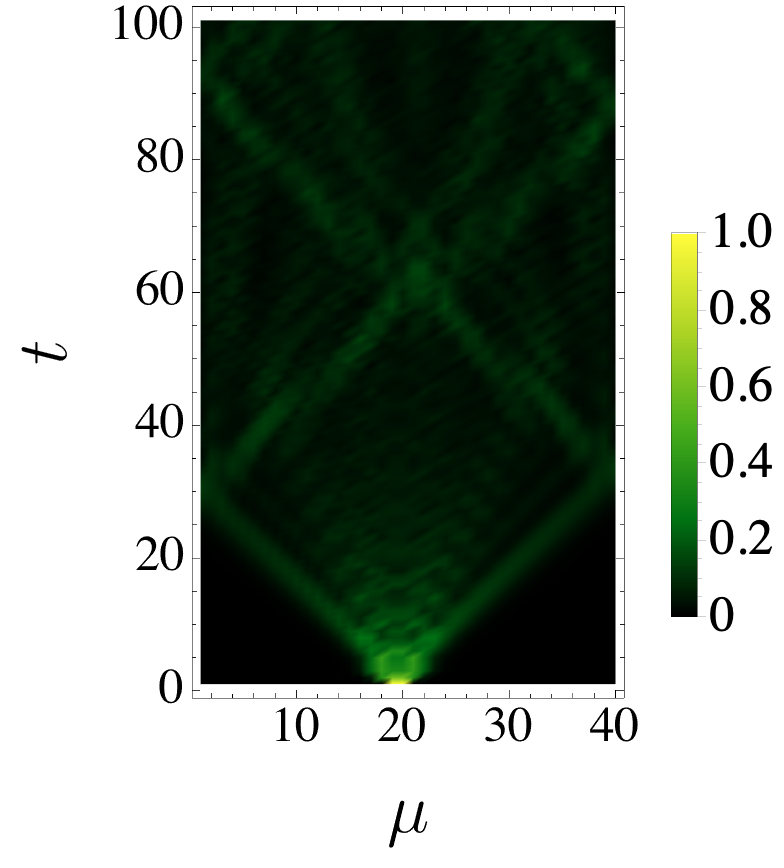}
	\label{fig_dw_close}}
 \subfigure[]{%
	\includegraphics[width=.4\textwidth,height=.27\textwidth]{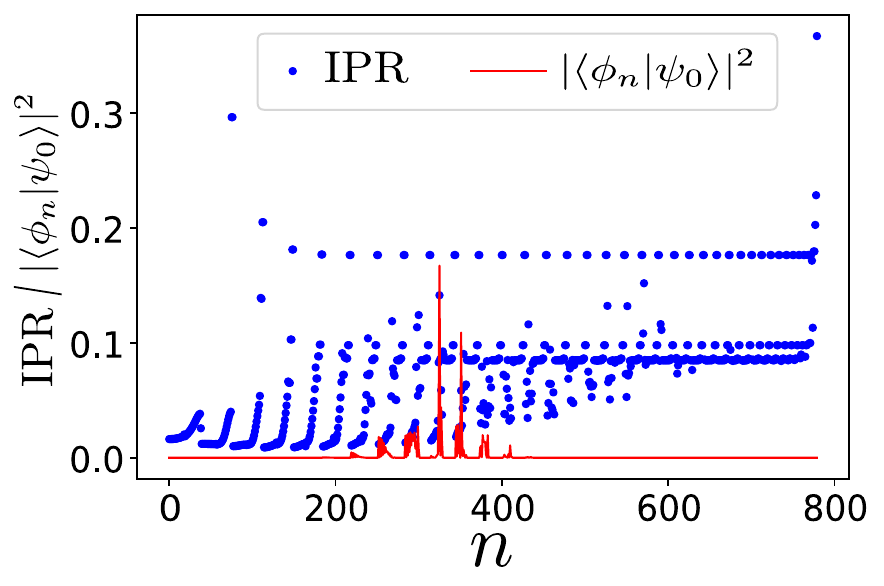}
	\label{fig_overlap_separate}}
 \subfigure[]{%
	\includegraphics[width=.4\textwidth,height=.27\textwidth]{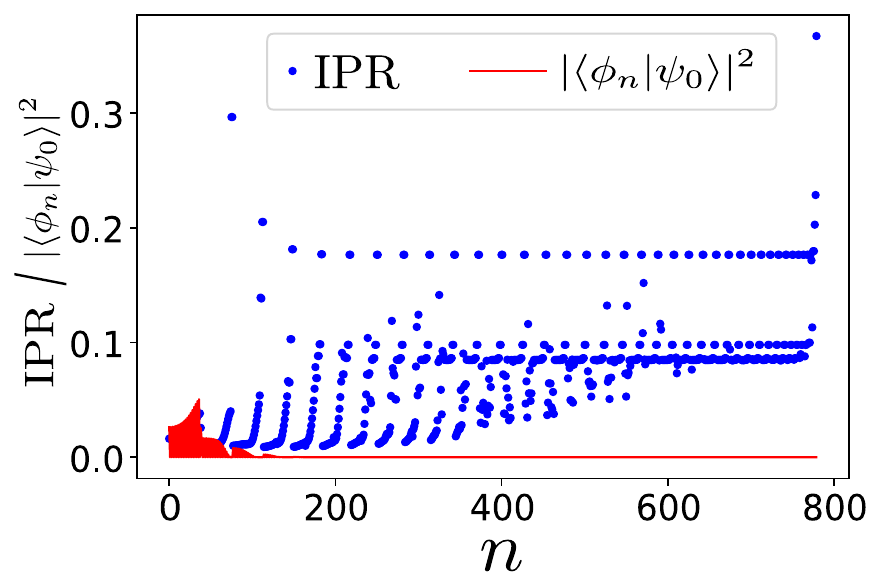}
	\label{fig_ovelap_close}}
\caption{ (Color online)  (Top) Time evolution of the number of fermions $\langle n_\mu (t)\rangle$ at site $\mu$ (domain wall position in the actual spin language) for the effective fermionic Hamiltonian $\tilde{H}$ obtained from the PXQ-QXP model. (a) Initial state with two fermions situated at separated sites $\mu=16$ and $\nu=25$, i.e.,  two separated domain walls.   We find localization of this state. (b) Initial state with two fermions at the adjacent sites $\mu=19$ and $\nu=20$, i.e., two nearby domain walls. We find delocalization of the state.  For both (a) and (b), we choose the total system size $N=40$, {$J=1$} and transverse field value $h=1.5$. (Bottom) The overlap of the two different initial states with the eigenstates of $\Tilde{H}$ (red). We also show the IPR value of all the eigenstates (blue). (c) Result for the initial state corresponding to (a). We find a significant overlap with eigenstates having high IPR.  (d) Result for the initial state corresponding to (b). We find a relatively large overlap with eigenstates having low IPR.}
\label{fig_pxq_qxp}
\end{figure*}
\subsection{Localization and delocalization dynamics in the PXQ–QXP model}
\label{sec_pxq-qxp-field}
Let us consider the case when a uniform longitudinal field $h$ is present in $H$ for the Hamiltonian constructed PXQ–QXP model to emphasize its difference with the PXQ model. In this scenario, the total Hamiltonian essentially corresponds to a quantum Ising chain with an additional longitudinal field,
\begin{equation}\label{eq_tfimlf}
    H'= - \frac{U}{4} \sum_{i=1}^{N-1}\sigma^z_i \sigma^z_{i+1} +\sum_{i=2}^{N-1}\left( J\sigma_i^x-\frac{h}{2}\sigma_i^z \right) + \frac{U}{4}\left(\sigma^z_1 - \sigma^z_N\right).
\end{equation}
This model has been studied previously in the strong Ising coupling limit ($U\gg J, h$) in the context of domain wall confinement~\cite{mazza19,james19,lerose20,collura22} and also in the context of Hilbert space fragmentation in higher dimensions~\cite{Yoshinaga22,Hart22}.
For one dimension, the effective dynamics will be essentially equivalent to the PXQ-QXP model.

Consider the first excited-energy subspace of the PXQ–QXP model, i.e., only one ``10" excitation is present in the bulk of the system. In this case, the system can be described by an equivalent Hamiltonian with two interacting quasi-particles corresponding to two domain walls ``01" (kink) and ``10" (anti-kink) present in the system~\cite{lerose20}. 
Then, the dynamics is described by the Hamiltonian $\tilde{H}$ with the matrix elements:
\begin{eqnarray}
    \braket{\mu\pm 1, \nu |\tilde{H}|\mu, \nu} &=& \braket{\mu, \nu\pm 1 |\tilde{H}|\mu, \nu}=J \\
    \braket{\mu,\nu|\tilde{H}|\mu,\nu} &=& -h |\mu-\nu| .
\end{eqnarray}
where $\ket{\mu,\nu}$ is the state with
$\mu$ and $\nu$ referring to the position of two quasi-particles.
Here, we neglect the constant energy term $(N-2)h/2$. 
This effective dynamics was previously discussed in Refs.~\cite{mazza19,james19,lerose20}.
In terms of the fermionic operators, the Hamiltonian in the two-particle sector assumes the following form,
\begin{equation}
\tilde{H}=J\sum_{\mu=1}^{N-2} \left( c_{\mu}^{\dagger} c_{\mu+1} + \mathrm{h.c.}\right) - \sum_{\mu, \nu =1}^{N-1} h |\mu - \nu| n_\mu n_\nu,
\end{equation}
where $n_\mu=c_\mu^\dagger c_\mu$ is the fermionic number operator for the $\mu$th site.

To see the dynamics in this model, we study the non-equilibrium evolution of the local fermionic number  $\langle n_\mu (t)\rangle= \bra{\psi(t)} c_\mu^{\dagger}c_\mu \ket{\psi(t)}$ for all $\mu$ as shown in the top panel of Fig.~\ref{fig_pxq_qxp}. In Fig.~\ref{fig_dw_separate}, we plot $\langle n_\mu (t)\rangle$ for the initial state in which two fermions are located at distant sites (i.e., for a large domain in the original spin chain where two domain walls are separated by a large distance), then each fermion (domain wall) remains localized at their initial position. This is because of the Wannier-Stark localization. As the distance between two fermions is large compared to the localization length $\xi$ for the Wannier-Stark ladder, they behave as isolated particles in the linear potential.  
On the other hand, if we consider an initial state where two fermions are situated at the adjacent sites (i.e., when two domain walls in the original spin chain are situated at adjacent bonds), the dynamics shows a different behavior (see Fig.~\ref{fig_dw_close}). In this case, as the distance between two fermions is comparable with the localization length $\xi$ for the Wannier-Stark ladder, we see that the quantity $\langle n_\mu (t)\rangle$ spreads over the entire system as time evolves, i.e., the system undergoes a delocalization dynamics. 
Note that this phenomenology was proposed in Ref.~\cite{lerose20}. 

To better understand the localization and delocalization phenomena for different initial states,
we further calculate the {inverse participation ratio (IPR)} for each eigenstate $\ket{\phi_n}$ of the Hamiltonian $\Tilde{H}$ as 
\begin{equation}
	\mathrm{IPR}(n)=\sum_F \lvert \bra{F}\phi_n\rangle\rvert^4,
\end{equation}
where $\{\ket{F}\}$ are the Fock space basis for two fermions.
The different dynamical behavior for the two different types of initial states can be understood by looking at the overlap of the two initial states with the eigenstates of the Hamiltonian $\Tilde{H}$.  We see that the localized initial states, in which two fermions are situated at distant sites, have significant overlap with the eigenstates having relatively high IPR values.
On the other hand, the delocalizing initial states, in which two fermions are situated at adjacent sites, have a relatively large overlap with the eigenstates that have very small IPR values. This is shown in Fig.~\ref{fig_overlap_separate} and Fig.~\ref{fig_ovelap_close}. Note that  the existence of atypical energy eigenstates exhibiting athermal features for the quantum Ising chain with an additional longitudinal field and strong Ising coupling were reported in Ref.~\cite{james19}.
Our results are consistent with their results and also clarify the mechanism of the initial-state dependence of the (de)localization dynamics by considering the overlap with the eigenstates.


\subsection{Additional Ising interactions}\label{Isingint}
{While we have considered a non-interacting Hamiltonian in Eq.~\eqref{eq_ham} so far, we can also start from an interacting Hamiltonian, such as the transverse field Ising model with a longitudinal field}
\begin{align}
H=\sum_{i=1}^{N-1}V\sigma_i^z\sigma_{i+1}^z+\sum_{i=2}^{N-1}\left(J \sigma_i^x -  \frac{h}{2} \sigma_i^z\right),
\end{align}
where $V$ is arbitrary.
Since the PXQ model and the PXQ-QXP model conserve the Ising energy, we can conclude that the existence of the Ising-interaction term does not change the above arguments at all.
In particular, if $V, J$, and $h$ are comparable, this model exhibits thermalization because of the ETH.
By adding strong dissipation, however, this model becomes integrable (for $h=0$) or localized (for $h\neq 0$).

\begin{figure}
\centering
\includegraphics[width=.38\textwidth,height=.28\textwidth]{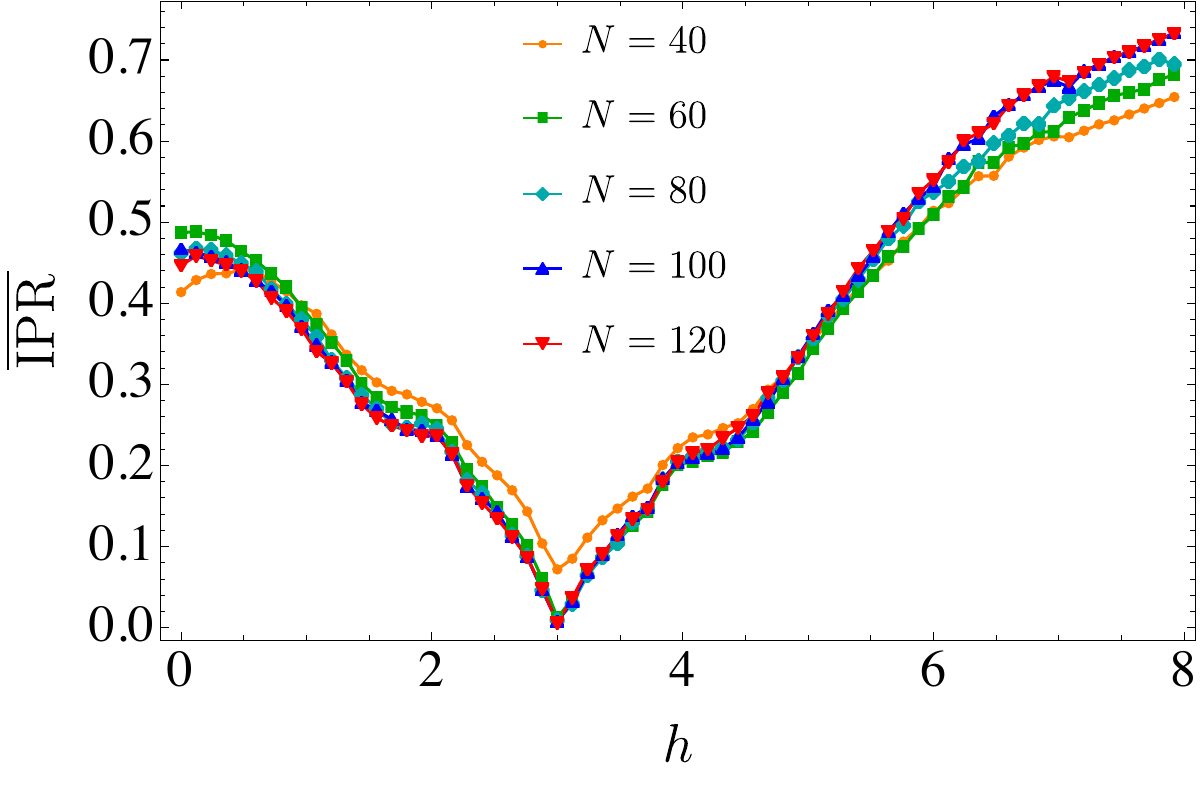}
\caption{ (Color online) Variation of the mean IPR for the eigenstates in a window $\Delta_E$ with $|\Delta_E|=60$ about the middle of the spectrum of $\Tilde{H}$ with respect to the longitudinal field $h$ for different system sizes $N=40, 60, 80, 100, 120$. Here, we choose the parameter {$J=1$}, $g=1.5$ and add a small disorder on-site potential of strength $[-10^{-4},10^{-4}]$ to break any underlying symmetries in the system. The variation of IPR for each curve shows a dip at the point $h=2g$. 
}
\label{fig_iprh_twopxq}
\end{figure}

\begin{figure}
\centering
\includegraphics[width=.38\textwidth,height=.28\textwidth]{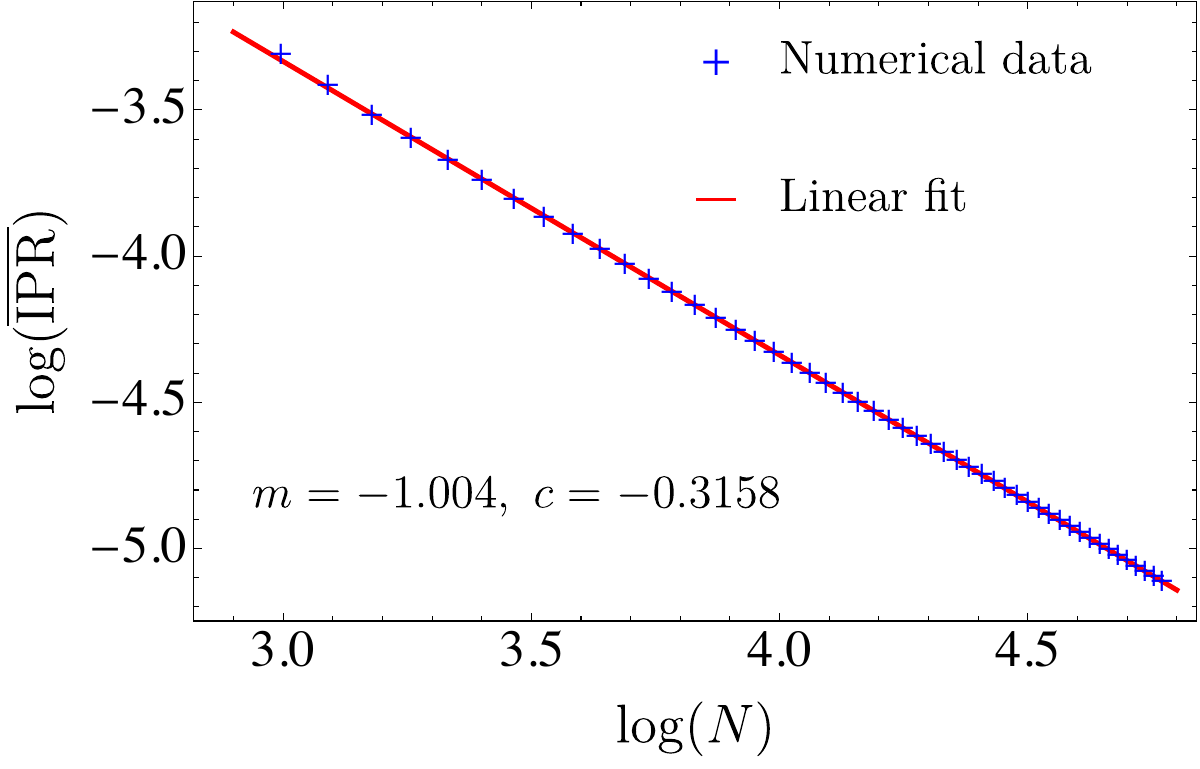}
\caption{ (Color online) Scaling for the mean IPR with the system size at the point $h=2g$.
We plot the mean IPR with the system size $N$ in logarithmic scales for the point $h=2g$ with {$J=1$} and $g=1.5$. The linear fitting $\log(\mathrm{\overline{IPR}})=m\log(N)+c$ with the slope $m\simeq-1$, indicates that the IPR for a typical eigenstate scales as $\mathrm{IPR} \sim 1/N$.
}
\label{fig_ipr_n_log}
\end{figure}

\section{Coupled PXQ chain}\label{sec_couple_pxq}
In Sec.~\ref{stark_localization}, we have shown that the PXQ model exhibits the single-particle Wannier-Stark localization in the presence of a non-zero magnetic field. In this section, we consider a more non-trivial model and localization phenomena by coupling two PXQ chains through the Ising interaction between spins of two chains. For this purpose, with our choice of jump operators $L_i=Q_i P_{i+1}$, we consider the unitary part in the Markovian dynamics in Eq.~\eqref{eq_gksl_ex} to be governed by the following Hamiltonian
\begin{eqnarray}
    H &=& \sum_{\alpha=1,2} \Bigg[\sum_{i=2}^{N-1} \Big( J\sigma_{i,{\alpha}}^x - \frac{h}{2} \sigma_{i,{\alpha}}^z \Big)  
    \non\\ &    &~~~~~~+\sum_{i=1}^{N-1}V\sigma_{i,{\alpha}}^z \sigma_{i+1,{\alpha} }^z\Bigg]+\sum_{i=1}^{N} g\sigma_{i,{1}}^z \sigma_{i,{2}}^z, 
\end{eqnarray}
where $\sigma^{x,y,z}_{i,\alpha}$ denotes the Pauli operator on the $i$th  site for the $\alpha$th chain ($\alpha=1,2$), and
$g$ and $V$ are the strengths of interchain and intrachain Ising interactions, respectively. We recall that the parameter $J$ is very small compared to the strength of dissipation $\gamma$. The presence of $\sigma^z_{i,{\alpha}}$ and $\sigma_{i,1}^z \sigma_{i,2}^z$ terms in $H$ does not change the structure of the emergent decoherence-free subspaces discussed in Sec.~\ref{sec_pxq}. Furthermore, as discussed in Sec.~\ref{Isingint}, the intra-chain Ising interaction term with strength $V$ can be dropped since it only gives the constant term in the subspaces of our interest.
Therefore, in the fermionic picture, the effective Hamiltonian to describe the dynamics in the zero frozen block sector of emergent decoherence-free subspaces assumes the following form,
\begin{eqnarray}
    \Tilde{H} = \sum_{\alpha=1,2} \Bigg[ \sum_{\mu=1}^{N-2}J\left(c_{\mu,{\alpha}}^{\dagger}c_{\mu+1,{\alpha}} + \mathrm{h.c.}\right) + \sum_{\mu=1}^{N-1}\mu h n_{\mu,{\alpha}} \Bigg] \non \\- \sum_{\mu_1=1}^{N-1} \sum_{\mu_2=1}^{N-1}2g\lvert \mu_1 - \mu_2\rvert n_{\mu_1,1}n_{\mu_2,2}+\mathrm{(constant)}, \non \\
    \label{eq_ham_coupledpxq}
\end{eqnarray}
where $n_{\mu,{\alpha}}=c_{\mu,{\alpha}}^{\dagger}c_{\mu,{\alpha}}$ is the fermionic number operator at the $\mu$th bond in the $\alpha$th chain. Therefore, in the fermionic model, each fermion experiences a linear potential along the chain and an interchain interaction proportional to the distance between two fermions. This scenario is equivalent to the two coupled PXQ chains as a ladder. For $h=0$, two fermions experience a confinement potential due to inter-chain interactions, and they  will remain localized by forming an inter-chain bound state, if two fermions are initially far apart. This is similar to the bound state formed by two domain walls in the case of a single quantum Ising chain with a non-zero longitudinal field, as discussed in Sec.~\ref{sec_pxq-qxp-field}.” However, the coupled PXQ chain shows an interesting delocalization phenomenon when we apply a non-zero longitudinal field, especially when $h=2g$, as we will discuss below.

We calculate the mean IPR for the eigenstates $\ket{\phi_n}$ of the above Hamiltonian as 
\begin{equation}
	\overline{\mathrm{IPR}}=\frac{1}{|\Delta_E|} \sum_{n \in \Delta_E} \sum_{F'} \lvert \bra{F'}\phi_n\rangle\rvert^4,
\end{equation}
where $\{\ket{F'}\}$ are the $(N-1)^2$ number of basis states for the two fermionic chains and the average is taken over a set of energy levels $\Delta_E$ at the middle of the energy spectrum ($|\Delta_E|$ denotes the number of energy levels for $\Delta_E$). 
In Fig.~\ref{fig_iprh_twopxq}, we show the variation of the mean IPR with increasing the longitudinal field $h$ for different system sizes. For each curve, the mean IPR shows a dip at the point $h=2g$. Furthermore, we investigate the scaling of the mean IPR with the system size. For the point $h=2g$, the mean IPR scales as $\textrm{IPR}\sim 1/N$ with the system size, which is manifested in Fig.~\ref{fig_ipr_n_log}.
This is in contrast with other values of $h$, where no such decay is observed.
Thus, the system is localized for all values of $h$ except the point $h=2g$, where it shows a partial delocalization behavior. 

\begin{figure}
\centering
\subfigure[]{
\includegraphics[width=.23\textwidth,height=.2\textwidth]{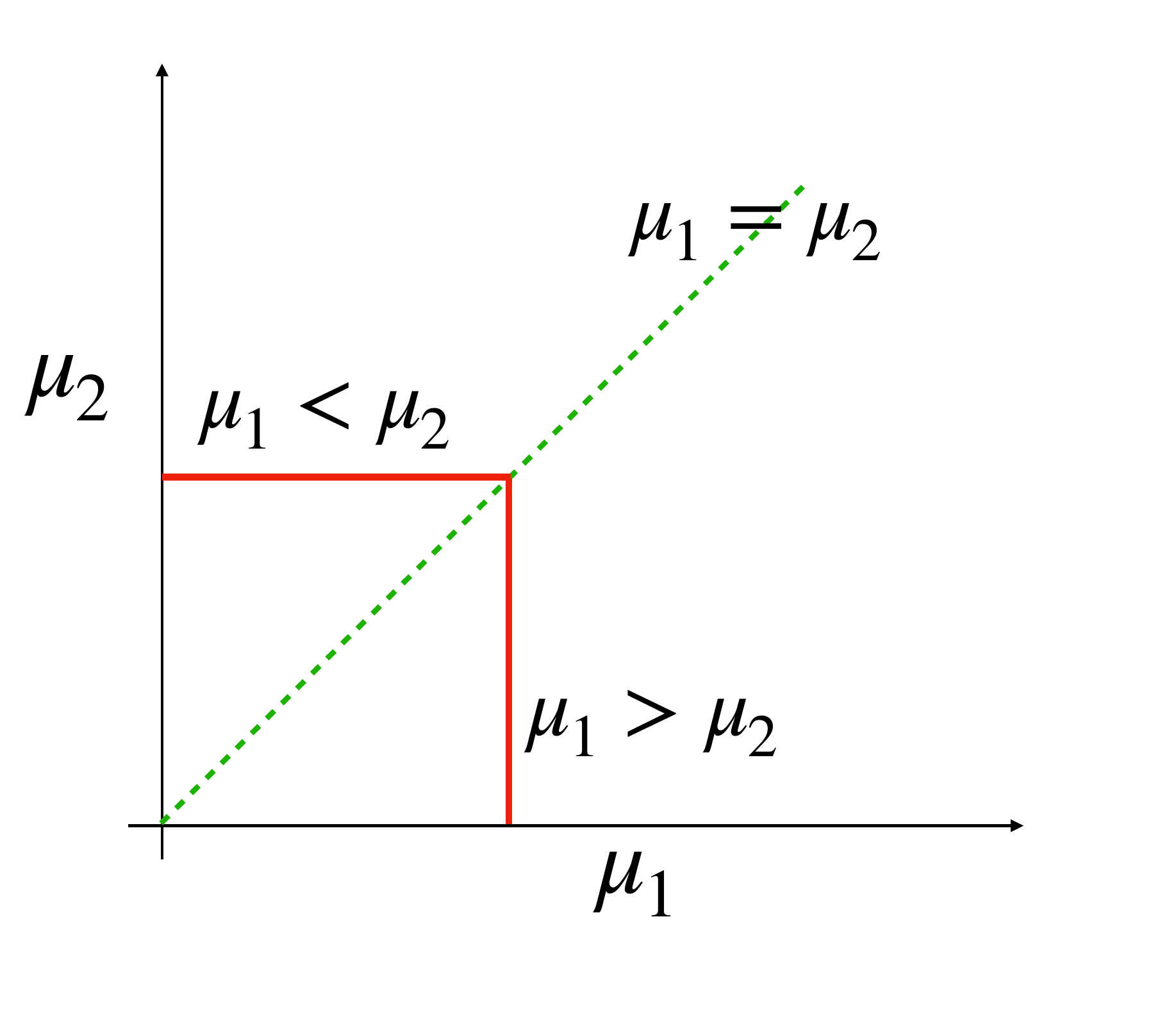}
        \label{fig_dis_diagram}}
\subfigure[]{%
\includegraphics[width=.225\textwidth,height=.2\textwidth]{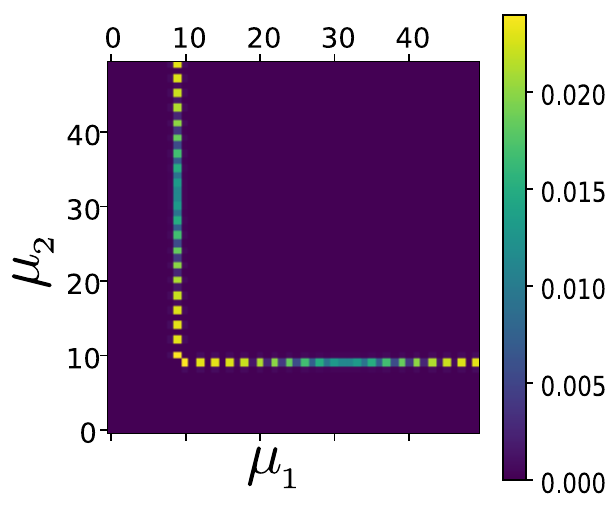}
	\label{fig_h2g}}
\subfigure[]{%
\includegraphics[width=.225\textwidth,height=.2\textwidth]{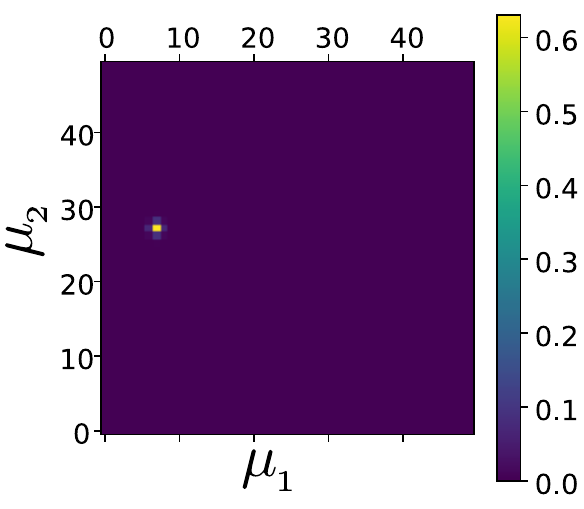}
	\label{fig_hless}}
 \subfigure[]{%
\includegraphics[width=.22\textwidth,height=.2\textwidth]{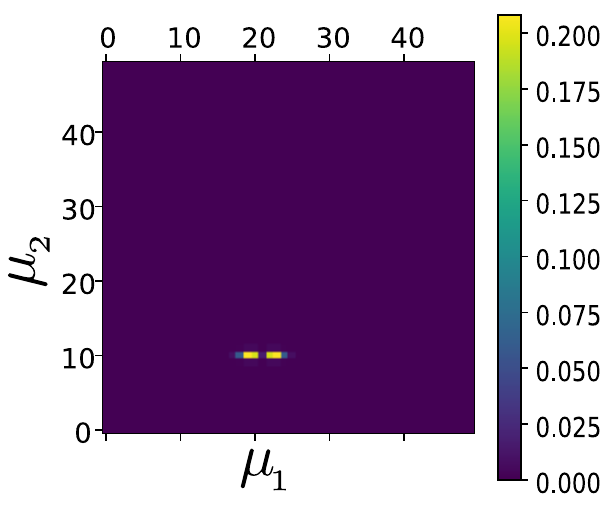}
	\label{fig_hgreater}}
\caption{ (Color online) (a) Diagram to illustrate the variation of the potential-energy term $\Xi(\mu_{1},\mu_{2})$ for the case $h=2g$. For the region $\mu_{1}\geq \mu_{2}$, the potential varies linearly in $\mu_{2}$ along the vertical red line. On the other region $\mu_{1}<\mu_{2}$, $\Xi(\mu_{1},\mu_{2})$ changes linearly with $\mu_{1}$ along the horizontal red line. 
(b-d) The distribution of a particular eigenstate $\ket{\phi_{n=860}}$ of $\tilde{H}$ over the lattice points on two chains of size $L=50$ for the longitudinal-field value (b) $h=2g$, (c) $h=0.1g$ and (d) $h=2.5g$. In all three figures, we choose {$J=1$}, $g=1.5$ and add a small disordered on-site potential of strength $[-10^{-4},10^{-4}]$ to break any underlying symmetries in the system. (b) The eigenstate is delocalized over the region where both fermions experience constant potential. In  cases (c) and (d), the fermions always experience linear potential, and thus the eigenstates are localized. 
}
\label{fig_egs_distribution}
\end{figure}

We can understand the partial delocalization by looking at the potential-energy terms in the Hamiltonian $\Tilde{H}$ in Eq.~\eqref{eq_ham_coupledpxq} when two fermions are located at $\mu_{1}$ and $\mu_{2}$ on the two chains:
\begin{equation}
    \Xi (\mu_{1},\mu_{2}) = \mu_{1}h + \mu_{2} h -2g|\mu_{1}-\mu_{2}|.
\end{equation}
For the case $h=2g$, the potential-energy term takes the following form
\begin{eqnarray}
    \Xi (\mu_{1},\mu_{2}) = \begin{cases}
2h\mu_{2},~~\text{for}~ \mu_{1} \geq \mu_{2},\\
2h\mu_{1},~~\text{for}~ \mu_{1} < \mu_{2}.
\end{cases}
\end{eqnarray}
The variation of $\Xi (\mu_{1},\mu_{2})$ is shown in Fig.~\ref{fig_dis_diagram}. For the region $\mu_{1} \geq \mu_{2}$, the potential linearly depends on $\mu_{2}$ only. Thus, for a fixed $\mu_{2}$, the other particle sees a constant potential and shows delocalization in this region over the chain $\alpha=1$. A similar delocalization happens for the particle in the chain  $\alpha=2$ in the region $\mu_{1} < \mu_{2}$. 
Consequently, the eigenstates with $h=2g$ have a unique structure: as shown in Fig.~\ref{fig_h2g}, the eigenstates $\ket{\phi_n}$ generally have a large overlap with states given by $\ket{\mu^{(n)}}_1\ket{\mu^{(n)}}_2, \ket{\mu^{(n)}}_1\ket{\mu^{(n)}+1}_2,\cdots,$ $ \ket{\mu^{(n)}}_1\ket{N-1}_2,  \ket{\mu^{(n)}+1}_1\ket{\mu^{(n)}}_2, \cdots, \ket{N-1}_1\ket{\mu^{(n)}}_2$, where $\mu^{(n)}$ is some point depending on $n$ and the subscripts denote the first and the second chains.
Importantly, these states amount to the number that typically increases as $\propto N$ (instead of localization $\propto N^0$ or the full delocalization $\propto N^2$), which is consistent with the scaling behavior of the IPR in Fig~\ref{fig_ipr_n_log}.

\section{Conclusions}
\label{sec_discussion}
In conclusion, our study introduces an approach to constructing kinetically constrained models through Markovian quantum dynamics subjected to strong dissipation, using the knowledge of decoherence-free subspaces. {We consider engineering the GKSL  master equation using classical noise, revealing the emergence of decoherence-free subspaces and enabling the realization of constrained quantum many-body unitary dynamics.} Importantly, our findings highlight the enhancement of dynamical constraints by the Markovian dynamics compared to those derived from the Hamiltonian one with strong interactions same as the jump operators.

Through a detailed exploration of a one-dimensional spin system with two-site jump operators, we exemplify the resulting kinetically constrained ``PXQ" model. Although this Lindbladian–constructed model exhibits a free domain–wall motion in the emergent decoherence-free subspace without frozen blocks,  it can show disconnected constrained dynamics for subspaces with frozen blocks. This constrained dynamics is different from its Hamiltonian–constructed counterpart. Furthermore, under the influence of a uniform magnetic field, the PXQ model exhibits domain-wall localization, reminiscent of the well-known Wannier-Stark localization, in each of the subsectors of the emergent decoherence-free subspaces. The dynamics of the corresponding Hamiltonian–constructed model in the ground-state subspace is exactly the same as the dynamics of the Lindbladian–constructed PXQ model in the zero-frozen-block sector. However, the dynamics is completely different in the two models in their higher energy subspaces and the sectors with the higher number of frozen blocks. 
Indeed, the Hamiltonian-constructed model under a longitudinal field shows intricate localization and delocalization behaviors of domain walls depending on the initial state, in contrast to the simple Wannier-Stark localization.

We further explore another type of model, which is introduced by coupling two PXQ chains in the presence of a magnetic field and inter-chain interactions. Remarkably, despite the presence of interactions, persistent domain-wall localization is observed in typical parameter regimes. Intriguingly, we reveal a non-trivial partial delocalization along a specific parameter line.

We point out that the paradigmatic PXP model can also be constructed using the Lindblad-construction method. For this, one needs to consider the two-site jump operators as $L_i=Q_i Q_{i+1}$ for a simple spin $1/2$ Hamiltonian (see Appendix.~\ref{app_pxp} for details). In this case, the dynamics in the emergent decoherence-free subspaces is the same as that in the Hamiltonian-constructed PXP model under
strong interaction between nearest–neighbor excitations
$Q_iQ_{i+1}$. This example of the PXP model indicates that the enhancement of constraints in the dynamics depends on the jump operators we choose. However, we convincingly demonstrate that, for general cases, the constraint is stronger in the case of the Lindbladian–constructed model compared to the Hamiltonian-constructed one. 

In essence, this work extends our understanding of kinetically constrained models in quantum dynamics as an emergent phenomenon in open quantum many-body systems, which offers insights into the interplay between dissipation and unitary dynamics. One challenging direction for future research from our results is the construction of models showing quantum many-body scars or the Hilbert space fragmentation that is hard to implement in previous approaches.

\begin{acknowledgments}
We are grateful to Yuta Sekino for valuable comments.
	We acknowledge the developers of QuSpin \cite{weinberg17,weinberg19}, with which we have carried out the numerical calculations.
This work was supported by JST ERATO Grant Number JPMJER2302, Japan.
\end{acknowledgments}

\appendix

\section{PXP Model}
\label{app_pxp}
In this section, we show that the well-known PXP model can also be constructed using the Lindblad dynamics if we consider the jump operators $L_i=Q_i Q_{i+1}$ and the Hamiltonian $H=J\sum_{j=2}^{N-1}\sigma^x_j$. With these choices of $H$ and $L_i$, and in large {dissipation strength} $\gamma$ limit, we recall the unperturbed  and perturbed parts in the Liouvillian (Eq.~\eqref{eq_gksl_ex}) as,
\begin{eqnarray}\label{eq_perturb_pxp}
	\mathcal{L}_0&=&\frac{\gamma}{2}\sum_{j=1}^{N-1} \Big( 2 Q_i Q_{i+1} \otimes Q_i Q_{i+1} - Q_i Q_{i+1} \otimes I \non \\ &&~~~~~~~~~~~~~~~~~~~~~~~~~~~ -I \otimes Q_i Q_{i+1}\Big),  \\
	\mathcal{L}_1&=&-i J\sum_{j=2}^{N-1} \left(\sigma_j^x\otimes I - I\otimes \sigma_j^x\right).
\end{eqnarray}
We perform a Schrieffer–Wolff~\cite{schrieffer66,bravyi11} version of degenerate perturbation theory in the limit $J/\gamma\rightarrow0$.

Let us explicitly write the stationary-state subspace for $\mathcal{L}_0$ in terms of the eigenstates of the jump operators. Consider a computational  basis $\ket{\vec{n}}\equiv\ket{n_1, n_2,\cdots n_L}$, where $n_i$ is either $0$ or $1$. Then, this state becomes the eigenstate for $L_i$ as
 \begin{equation}
L_i \ket{\vec{n}}=f_i \ket{\vec{n}}.
 \end{equation}
There are two eigenvalues for each single jump operator $L_i$ with eigenvalue $f_i=0$ (when $Q_i Q_{i+1}=0$) and $f_i=1$ (when $Q_i Q_{i+1}=1$).  
Any state in the stationary state subspace for $\mathcal{L}_0$ can be written as a linear combination of $\ket{\vec{n}}\otimes\ket{\vec{n}'}$, where  $\ket{\vec{n}}$ and $\ket{\vec{n}'}$ are the eigenstates of $\{L_i\}$ with the same $\{f_i\}_{i=1}^{L-1}$. Considering $i$-th and $(i+1)$-th sites, the eigenstates of $L_i$ are given by $\ket{00}, \ket{01},\ket{10}$ and $\ket{11}$. The first three eigenstates are degenerate with eigenvalues $f_i=0$, and the eigenvalue corresponding to the eigenstate $\ket{11}$ is $f_i=1$. Thus, the stationary-state subspace for $\mathcal{L}_0$ (which leads to the emergent decoherence-free subspaces for $\mathcal{L}$) is spanned by the following states:
\begin{eqnarray}\label{eds_pxp}
	\Big\{\ket{00}\otimes\ket{00}, \ket{00}\otimes\ket{01}, \ket{00}\otimes\ket{10},  \ket{01}\otimes\ket{00}, \non\\  \ket{01}\otimes\ket{01}, \ket{01}\otimes\ket{10},  \ket{10}\otimes\ket{00}, \ket{10}\otimes\ket{01}, \non\\ \ket{10}\otimes\ket{10}, ~\text{and}~\ket{11}\otimes\ket{11}\Big\}. 
\end{eqnarray}
Therefore, in the first-order perturbation term for the effective Liouvillian, the only allowed transitions in the stationary-state subspace are:
\begin{eqnarray}
	\ket{000}\otimes\ket{\cdot \cdot \cdot} \longleftrightarrow \ket{010}\otimes\ket{\cdot \cdot \cdot }, \\
	\ket{\cdot \cdot \cdot}\otimes\ket{000} \longleftrightarrow \ket{\cdot \cdot \cdot}\otimes\ket{010}.
\end{eqnarray}
Furthermore, the configuration $\ket{11}\otimes\ket{11}$ forms a frozen block in this case. Then, we can express the effective Hamiltonian in the form of the PXP Hamiltonian
\begin{equation}
	H_{\mathrm{eff}}=J\sum_{i=2}^{N-1} P_{i-1}\sigma_i^x P_{i+1},
\end{equation}
where we choose open boundary condition.

Similar to the PXQ model, the emergent decoherence-free subspaces become decomposed into block diagonal sectors depending upon the number of frozen blocks present in the system. Furthermore, each sector is split into subsectors depending on the position of the frozen blocks. 
However, in this case, the dynamics in the emergent decoherence-free subspaces is exactly the same as the dynamics of the Hamiltonian-constructed PXP model under strong interaction between nearest–neighbor excitations $Q_i Q_{i+1}$. 
The different energy subspaces (corresponding to the number of excitations of adjacent spins) of the Hamiltonian-constructed PXP model are exactly equivalent to the different sectors (corresponding to the number of frozen blocks) of the emergent decoherence-free subspaces of the Lindbladian-constructed PXP model.

\section{Second-order perturbation theory}\label{app_secondorder}
Here, we calculate the second-order terms in the perturbation theory for the PXP model case (as discussed in Sec.~\ref{app_pxp}). However, the calculation can be easily generalized to the other choices of jump operators discussed in this paper. 

To the second-order term in $J/\gamma$, the Schrieffer–Wolff transformation gives the effective Liouvillian as
\begin{equation}
	\mathcal{L}_{\mathrm{eff}}^{\left(2\right)}=-\mathcal{P}\mathcal{L}_1\left(1-\mathcal{P}\right)\mathcal{L}_0^{-1}\left(1-\mathcal{P}\right)\mathcal{L}_1\mathcal{P}.
	\label{second_perturb}
\end{equation} 
The action of the projection operator $\mathcal{P}$ first chooses a state from the stationary state subspace for $\mathcal{L}_0$ given in Eq.~\eqref{eds_pxp}. Now, the action of $\mathcal{L}_1$ (see Eq.~\eqref{eq_perturb_pxp}) on the stationary state $\ket{\vec{n}}\otimes \ket{\vec{n}'}$ is given by
\begin{equation}
	\mathcal{L}_1 \ket{\vec{n}}\otimes \ket{\vec{n}'}=-iJ \sum_j \left( \ket{\vec{n}_{\left(j\right)}}\otimes\ket{\vec{n}'}-\ket{\vec{n}}\otimes\ket{\vec{n}'_{\left(j\right)}}\right),
\end{equation}
where $\ket{\vec{n}_{\left(j\right)}}=\sigma_j^x\ket{\vec{n}}$ is the state where the $j$th spin is flipped in $\ket{\vec{n}}$. Now, the state $\ket{\vec{n}_{\left(j\right)}}$ ($\ket{\vec{n}_{\left(j\right)}'}$) can have eigenvalues of $\{L_i\}$ that are different from those for $\ket{\vec{n}'}$ ($\ket{\vec{n}}$). 
Thus, the resulting state $\mathcal{L}_1\ket{\vec{n}}\otimes \ket{\vec{n}'}$ may no longer be in the stationary-state subspace for $\mathcal{L}_0$. 
Furthermore, the projection to the subspace $(1-\mathcal{P})$ determines that the resulting state $(1-\mathcal{P})\mathcal{L}_1\ket{\vec{n}}\otimes \ket{\vec{n}'}$ is  not in the stationary-state subspace. Therefore, we have
\begin{equation}
	(1-\mathcal{P})\mathcal{L}_1 \ket{\vec{n}}\otimes \ket{\vec{n}'}=-iJ {\sum_j}' \left( \ket{\vec{n}_{\left(j\right)}}\otimes\ket{\vec{n}'}-\ket{\vec{n}}\otimes\ket{\vec{n}'_{\left(j\right)}}\right),
\end{equation}
where ${\sum}_j'$ indicates  that we take the sum over $j$ such that $\ket{\vec{n}_{\left(j\right)}}$ and $\ket{\vec{n}'}$ (also $\ket{\vec{n}}$ and $\ket{\vec{n}'_{\left(j\right)}}$) correspond to eigenstates of $\{L_i\}$ with different eigenvalues $\{f_i\}_{i=1}^{L-1}$. We note that the flipping of a single spin at $j$th position in $\ket{\vec{n}}$ or $\ket{\vec{n'}}$ can only change the $(j-1)$th eigenvalue $f_{j-1}$ or the $j$th eigenvalue $f_j$ or simultaneously both the eigenvalues $f_{j-1}$ and $f_{j}$  in the set $\{f_i\}_{i=1}^{L-1}$.

In the next step, the action of $\mathcal{L}_0^{-1}$ on $(1-\mathcal{P})\mathcal{L}_1\mathcal{P}\ket{\vec{n}}\otimes \ket{\vec{n}'}$ does not change the state, rather it has an effect such that $\mathcal{L}_0^{-1} \ket{\vec{n}_{\left(j\right)}}\otimes\ket{\vec{n}'} = \lambda ^{-1} \ket{\vec{n}_{\left(j\right)}}\otimes\ket{\vec{n}'}$ with
\begin{eqnarray}
	\frac{2\lambda}{\gamma}&=&\sum_{k=1}^{N-1} \Big[ 2(n_{(j),k} n_{(j),k+1} )(n'_k n'_{k+1})  \\ && ~~~~~~~~~~~~~~~~ - n_{(j),k} n_{(j),k+1} -n'_k n'_{k+1}\Big]. \non
\end{eqnarray}
The right-hand side of the above equation can take values either $-1$ or $-2$ depending on the configuration in $\ket{\vec{n}_{\left(j\right)}}$. One can check that $2\lambda/\gamma$ will take value $-1$ ($-2$) if $\ket{\vec{n}_{\left(j\right)}}=\sigma_j^x\ket{\vec{n}}$ changes only one eigenvalue  $f_{j-1}$ or $f_{j}$ (two eigenvalues $f_{j-1}$ and $f_{j}$) in  $\{f_i\}_{i=1}^{L-1}$.

As all the terms in $\mathcal{L}_0^{-1}(1-\mathcal{P})\mathcal{L}_1\mathcal{P}\ket{\vec{n}}\otimes \ket{\vec{n}'}$ are already in the subspace $(1-\mathcal{P})$, further projection to the space $(1-\mathcal{P})$ does not change the state. 
Now the action of $\mathcal{L}_1$ on $\left(1-\mathcal{P}\right)\mathcal{L}_0^{-1}\left(1-\mathcal{P}\right)\mathcal{L}_1\mathcal{P}$ is given by
\begin{widetext}
	\begin{equation}
		\mathcal{L}_1\left(1-\mathcal{P}\right)\mathcal{L}_0^{-1}\left(1-\mathcal{P}\right)\mathcal{L}_1\ket{\vec{n}}\otimes\ket{\vec{n}'}=-\frac{J^2}{\lambda}\sum_{k}{\sum_{j}}' \left( \sigma_k^x \ket{\vec{n}_{\left(j\right)}}\otimes\ket{\vec{n}'} - \sigma_k^x\ket{\vec{n}}\otimes\ket{\vec{n}'_{\left(j\right)}} -\ket{\vec{n}_{\left(j\right)}}\otimes\sigma_k^x \ket{\vec{n}'}+\ket{\vec{n}}\otimes\sigma_k^x \ket{\vec{n}'_{\left(j\right)}}\right).
	\end{equation}
\end{widetext}
Remember $\ket{\vec{n}_{\left(j\right)}}$ and $\ket{\vec{n}'}$ (also $\ket{\vec{n}}$ and $\ket{\vec{n}'_{\left(j\right)}}$) in the above expression have different sets of eigenvalues $\{f_i\}_{i=1}^{L-1}$.
Projecting back the above quantity to the stationary-state subspace of $\mathcal{L}_0$ with $\mathcal{P}$, one can find the matrix elements of the second-order effective Liouvillian as
\begin{widetext}
\begin{equation}
	\left[\mathcal{L}_{\mathrm{eff}}^{\left(2\right)}\right]_{\vec{m},\vec{m}',\vec{n},\vec{n}'}\sim-\frac{J^2}{\gamma} \sum_{k}{\sum_{j}}' \bra{\vec{m}}\otimes\bra{\vec{m}'}\left( \sigma_k^x \sigma_j^x \otimes I -\sigma_k^x\otimes\sigma_j^x - \sigma_j^x\otimes\sigma_k^x + I\otimes \sigma_k^x \sigma_j^x\right) \ket{\vec{n}}\otimes\ket{\vec{n}'},
	\label{eq_leff_2}
\end{equation}
\end{widetext}
where $\ket{\vec{m}}$ and $\ket{\vec{m}'}$ have the same eigenstates $\{f_i\}_{i=1}^{L-1}$. In other words, both $\ket{\vec{n}}\otimes\ket{\vec{n}'}$ and $\ket{\vec{m}}\otimes\ket{\vec{m}'}$ are in the stationary-state subspace for $\mathcal{L}_0$.
Here, we have used the fact that $\lambda$ is proportional to {dissipation strength} $\gamma$.

Looking at the allowed configurations in this  subspace, we can easily see that
the second-order effective Liouvillian $\mathcal{L}_{\mathrm{eff}}^{\left(2\right)}$ will have non-zero matrix elements only if the allowed values of $k$ are $j-1$, $j$ and $j+1$.  Note that all the coefficients in Eq.~\eqref{eq_leff_2} are of the order of $J^2/\gamma$. Therefore, for time $t \ll \gamma/J^2$, it is natural to assume that the first-order effective Liouvillian can describe the dynamics of the system~\footnote{To be precise, because of the sum over different sites, the magnitude for $\mathcal{L}_\mathrm{eff}^{(2)}$ can be order of $J^2N/\gamma$. Indeed, we can rigorously show that the first-order perturbation is accurate when $t\ll \gamma/(JN)^2$ if $\gamma\gg JN$~\cite{Gong20_pra}. While this condition breaks for the thermodynamic limit $N\rightarrow\infty$, we conjecture that the first-order perturbation is accurate even for $N$-independent timescales if $\gamma\gg J$ for locally interacting systems and observables, as is the case for the unitary dynamics~\cite{Gong20_pra}.}. 

\bibliography{references_lhsf.bib}

\end{document}